\documentclass[a4paper,11pt,twoside]{article}

\pdfoutput=1

\makeatletter

\usepackage{amsmath,amssymb}

\usepackage[english]{babel}

\usepackage{cancel}
\usepackage{color}

\usepackage{graphicx}
\usepackage{geometry}

\usepackage{hyperref}

\usepackage[amsmath,hyperref,thmmarks]{ntheorem}

\usepackage{url}

\usepackage{enumerate}

\numberwithin{equation}{section}
\allowdisplaybreaks[1]

\newcommand{\be}{\begin{equation}}
\newcommand{\ee}{\end{equation}}

\newcommand\bbone{{ \mathbb{I}}}

\DeclareMathOperator{\tr}{Tr}

\newcommand\ham{{\text{Ham}}}

\newcommand{\labitem}[2]{\def\@itemlabel{\textbf{#1}}\item\def\@currentlabel{#1}\label{#2}}

\theoremsymbol{}
\theorembodyfont{\slshape}
\theoremheaderfont{\normalfont\bfseries}
\theoremseparator{}

\theorembodyfont{\upshape}
\theoremsymbol{\ensuremath{\blacklozenge}}

\theoremstyle{nonumberplain}
\theoremheaderfont{\scshape}
\theorembodyfont{\normalfont}
\theoremsymbol{\ensuremath{\blacksquare}}

\qedsymbol{\ensuremath{_\blacksquare}}
\theoremclass{LaTeX}

\newcommand{\institute}[1]{\newcommand{\@institute}{#1}}

\renewcommand{\maketitle}{
\vspace*{0.5\baselineskip}
{
\center\LARGE\noindent\@title\par
}%
\vspace{1.5\baselineskip}
{
\center\normalsize\noindent\ignorespaces\@author\par
}%
\vspace{0.5\baselineskip}
{
\center\normalsize\ignorespaces\@institute\par
}%
\vspace{2\baselineskip}
}%

\definecolor{hypercolor}{rgb}{0.1,0.2,0.6}

\hypersetup{     
unicode=false,      
pdftoolbar=true,    
pdfmenubar=true,    
pdffitwindow=true,  
pdfstartview={FitH},
pdftitle={Draft},    
pdfauthor={Antoine G\'er\'e, Tajron Juri\'c, Jean-Christophe Wallet},     
pdfsubject={Mathematical Physics},   
pdfcreator={LaTeX},  
pdfproducer={pdfTex},
pdfkeywords={Noncommutative Field Theory},  
pdfnewwindow=true,  
}  
\renewenvironment{thebibliography}[1]{%
\section*{References}%
\frenchspacing\small%
\begin{list}{[\arabic{enumi}]}%
{%
\usecounter{enumi}\parsep=2pt\topsep 0pt%
\settowidth{\labelwidth}{[#1]}%
\leftmargin=\labelwidth\advance\leftmargin\labelsep%
\rightmargin=0pt\itemsep=1pt\sloppy%
}%
}{\end{list}}

\begin{document}

\title{Noncommutative gauge theories on $\mathbb{R}^3_\lambda$: \\
Perturbatively finite models}

\author{Antoine G\'er\'e$^a$, Tajron Juri\'c$^b$, Jean-Christophe Wallet$^c$}

\institute{%
\textit{$^a$Dipartimento di Matematica, Universit\`a di Genova\\
Via Dodecaneso, 35, I-16146 Genova, Italy}\\
e-mail:\href{mailto:gere@dima.unige.it}{\texttt{gere@dima.unige.it}}\\[1ex]%
\textit{$^b$Ru\dj er Bo\v{s}kovi\'c Institute, Theoretical Physics Division\\
Bijeni\v{c}ka c.54, HR-10002 Zagreb, Croatia}\\
e-mail:\href{mailto:tjuric@irb.hr}{\texttt{tjuric@irb.hr}}\\[1ex]%
\textit{$^c$Laboratoire de Physique Th\'eorique, B\^at.\ 210\\
CNRS and Universit\'e Paris-Sud 11,  91405 Orsay Cedex, France}\\
e-mail:\href{mailto:jean-christophe.wallet@th.u-psud.fr}{\texttt{jean-christophe.wallet@th.u-psud.fr}}\\[1ex]%
}%

\date{\today}

\maketitle

\begin{abstract} 
We show that natural noncommutative gauge theory models on $\mathbb{R}^3_\lambda$ can accommodate gauge invariant harmonic terms, thanks to the existence of a relationship between the center of $\mathbb{R}^3_\lambda$ and the components of the gauge invariant 1-form canonical connection. This latter object shows up naturally within the present noncommutative differential calculus. Restricting ourselves to positive actions with covariant coordinates as field variables, a suitable gauge-fixing leads to a family of matrix models with quartic interactions and kinetic operators with compact resolvent. Their perturbative behavior is then studied. We first compute the 2-point and 4-point functions at the one-loop order within a subfamily of these matrix models for which the interactions have a symmetric form. We find that the corresponding contributions are finite. We then extend this result to arbitrary order. We find that the amplitudes of the ribbon diagrams for the models of this subfamily are finite to all orders in perturbation. This result extends finally to any of the models of the whole family of matrix models obtained from the above gauge-fixing. The origin of this result is discussed. Finally, the existence of a particular model related to integrable hierarchies is indicated, for which the partition function is expressible as a product of ratios of determinants.

\end{abstract}

\newpage

\section{Introduction}

Noncommutative Geometry (NCG) \cite{Connes} may provide an appealing way to overcome physical obstructions to the existence of continuous space-time and commuting coordinates at the Planck scale \cite{Doplich1}, triggering a new impulse in the studies on noncommutative field theories (NCFT). Actually, they appeared in their modern formulation a long time ago within String Field Theory \cite{witt1}. This was followed by models on the fuzzy sphere \cite{gm90}, gauge theories on almost commutative geometries \cite{mdv1} (for a review on fuzzy sphere and related see e.g \cite{madorebook}). NCFT on noncommutative Moyal spaces received a lot of attention from the end of the 90's, in particular from the viewpoint of perturbative properties and renormalisability \cite{minwala, Chepelev}. For reviews, see for instance \cite{dnsw-rev}.\par 

Progresses have been made in the area of NCFT on Moyal spaces $\mathbb{R}^n_\theta$, $n=1,2$ leading to perturbatively renormalisable scalar fields theories. These encompass the scalar $\phi^4$ model with harmonic term on $\mathbb{R}^2_\theta$ or $\mathbb{R}^4_\theta$ \cite{Grosse:2003aj-pc}, this latter being likely non-perturbatively solvable \cite{harald-raimar}, the translational and rotational invariant related $\phi^4$ models \cite{branqui-1}, \cite{adg-w111} together with fermionic versions \cite{vtw} and solvable models inherited from the LSZ model \cite{LSZ}. The situation for the gauge theories is not so favorable. Although the construction of gauge invariant classical actions can be easily done from suitable noncommutative differential calculi \cite{mdv88-99, cgmw-20}, the study of quantum properties is rendered difficult by technical complications stemming mainly from gauge invariance that supplement the UV/IR mixing problem inherent in NCFT on Moyal spaces. So far, the construction of a renormalisable gauge theory on $\mathbb{R}^4_\theta$ has not been achieved. On Moyal spaces, gauge invariant straightforward generalizations of the above harmonic term do not exist. In this respect, attempts to reconcile the features of the $\phi^4$ model with harmonic term with a gauge theoretic framework gave rise to the gauge invariant model obtained in \cite{Wallet:2007c}. Interestingly, this action can be interpreted as (related to) the spectral action of a particular spectral triple \cite{Grosse:2007jy} whose relationship to the Moyal geometry has been analysed in \cite{Wallet:2011aa}. Unfortunately, its complicated vacuum structure explored in \cite{GWW2} forbids the use of any standard perturbative treatment{\footnote{This technical obstruction can be circumvented on $\mathbb{R}^2_\theta$ for particular vacuum configurations \cite{MVW13}.}}. Alternative based
on the implementation of a IR damping mechanism have been proposed and studied \cite{blaschk1}, \cite{Blaschke:2009c}, \cite{bgw-13}. Although this damping mechanism is appealing, it is not known if it can produce a renormalisable gauge theory on $\mathbb{R}^4_\theta$. Besides, interpreting the action within the framework of some noncommutative differential geometry is unclear. Another appealing approach is the matrix model formulation of noncommutative gauge theory, initiated a long ago in \cite{matrix1}. For recent reviews, see \cite{matrix2}, \cite{matrix3}. This approach may in some cases allow one to go beyond the perturbative approach \cite{matrix4}. One interesting outcome is that it may provide a interpretation for the UV/IR mixing for some noncommutative gauge theories in terms of an induced gravity action. See e.g \cite{matrix5}.\par

Recently, scalar field theories on the noncommutative space $\mathbb{R}^3_\lambda$, a deformation of $\mathbb{R}^3$ preserving rotation invariance, have been studied in \cite{vit-wal-12}. These appear  to have a mild perturbative behavior and are (very likely) free of ultraviolet/infrared (UV/IR) mixing. In this respect, one may expect a more favorable situation for the gauge theories on $\mathbb{R}^3_\lambda$ than for those on $\mathbb{R}^4_\theta$. The space $\mathbb{R}^3_\lambda$, which may by viewed as a subalgebra of $\mathbb{R}^4_\theta$,  has been first introduced in \cite{Hammaa} and generalized in \cite{selene}. The use of the canonical matrix base introduced in \cite{vit-wal-12} (see also \cite{duflo}) renders the computation tractable, avoiding the complexity of a direct calculation in coordinates space. A first exploration of gauge theories on $\mathbb{R}^3_\lambda$ has been performed in \cite{gervitwal-13}, focused on a particular class of theories for which the gauge-fixed propagator can be explicitly computed rendering possible a one-loop analysis. The impact of the expected mild perturbative behavior of the loop diagrams was however tempered by the occurrence of a nonzero one-loop tadpole signaling quantum instability of the chosen vacuum. While further study of this quantum instability may reveal new interesting properties, it seemed desirable to undertake a more systematic investigation around the construction of other families of gauge theories on $\mathbb{R}^3_\lambda$ with stable vacuum and non trivial dynamics. Reconciling these two features seems to be out of reach in the case of Moyal spaces but can be achieved when dealing with $\mathbb{R}^3_\lambda$ .\par
In this paper, we show that natural noncommutative gauge theory models on $\mathbb{R}^3_\lambda$ can support gauge invariant harmonic terms, unlike the case of Moyal spaces. This stems from the existence of a relationship between the center of $\mathbb{R}^3_\lambda$ and the components of the gauge invariant 1-form canonical connection which arises in the derivation-based differential calculus underlying our construction. We focus our analysis on a family of (positive) gauge invariant actions whose field variables are assumed to be the covariant coordinates, i.e. the natural objects related to the canonical connection. Then, a suitable BRST gauge-fixing in the spirit of \cite{MVW13, stor-wal} gives rise to a family of matrix models with quartic interactions and kinetic operators (having compact resolvent). Their perturbative behavior is then examined. We first consider a subfamily of these matrix models for which interactions and kinetic operators leads to slight technical simplifications and compute the corresponding 2-point and 4-point functions at the one-loop order. We find that the respective contributions are finite. We then extend this result to arbitrary order and find that the amplitudes of the ribbon diagrams for the models pertaining to this subfamily are finite to all orders in perturbation. It appears that this perturbative finitude results from the conjunction a sufficient rapid decay for the propagator, the role played by the radius of the fuzzy sphere components of $\mathbb{R}^3_\lambda$ acting as a kind of cut-off together with the existence of an upper bound for the (positive) propagator depending only of the cut-off. We then extend this result to any of the matrix models of the whole family obtained from the above gauge-fixing. Finally, we point out the existence of a particular model related to integrable (2-d Toda) hierarchies and give the expression of the partition function as a product of ratios of determinants.\par

The paper is organized as follows. In section \ref{section2}, we present and discuss the construction of the relevant family of gauge invariant models. Useful properties on the (derivation based) noncommutative differential calculus together with the notion of noncommutative connection inherited from the (commutative) notion of Koszul connection are also recalled. Section \ref{section3} is devoted to the gauge-fixing and the perturbative analysis with the one-loop computations collected in the subsection \ref{subsection322} while subsection \ref{subsection33} deals with the finitude to arbitrary orders. In section \ref{section4}, we discuss the results and finally consider also a particular model for which the partition function can be related to ratios of determinants signaling a relation to integrable hierarchies.

\section{\texorpdfstring{Noncommutative gauge theories on $\mathbb{R}^3_\lambda$}{Noncommutative gauge theories}} \label{section2}

\subsection{\texorpdfstring{Basic properties of $\mathbb{R}^3_\lambda$}{R3l}}\label{subsection21}

The algebra $\mathbb{R}^3_\lambda$ has been first introduced in \cite{Hammaa} and further considered in various works \cite{selene,vit-wal-12,gervitwal-13}. Besides, a characterization of a natural basis has been given in \cite{vit-wal-12}. We refer to these references for more details. Here{\footnote{To simplify the notations, the associative $\star$-product for $\mathbb{R}^3_\lambda$ is understood everywhere in any product of elements of the algebra. Besides, summation over repeated indices is understood everywhere, unless explicitly stated.}}, it will be convenient to view $\mathbb{R}^3_\lambda$ as \cite{vit-wal-12,gervitwal-13}%
\begin{equation}
\mathbb{R}^3_\lambda=\mathbb{C}\left[x_1,x_2,x_3,x_0\right]/{\mathcal{I}}[{\mathcal{R}}_1,{\cal{R}}_2] \ , \label{defining} 
\end{equation}
where $\mathbb{C}\left[x_1,x_2,x_3,x_0\right]$ is the free algebra generated by the 4 (hermitean) elements (coordinates) $\left\{x_{\mu=1,2,3},\ x_0\right\}$ and ${\mathcal{I}}\left[{\mathcal{R}}_1,{\cal{R}}_2\right]$ is the two-sided ideal generated by the relations%
\begin{equation}
{\mathcal{R}}_1: \ [x_\mu,x_\nu] = i \lambda \varepsilon_{\mu\nu\rho} x_\rho \ , \quad
{\mathcal{R}}_2: \ x_0^2 + \lambda x_0 = \sum_{\mu=1}^3 x_\mu^2,\ \forall \mu,\nu,\rho=1,2,3\label{relat1}
\end{equation}
with $\lambda\ne0$. $\mathbb{R}^3_\lambda$ is a unital $*$-algebra, with complex conjugation as involution and center ${\cal{Z}}(\mathbb{R}^3_\lambda)$ generated by $x_0$ and satisfying the following strict inclusion $\mathbb{R}^3_\lambda\supsetneq U(\mathfrak{su}(2))$, where $U(\mathfrak{su}(2))$ is the universal enveloping algebra of the Lie algebra ${\mathfrak{su}}(2)$. Alternative (equivalent) presentations can be found in e.g \cite{selene, vit-wal-12, gervitwal-13}.\par%
As shown in \cite{vit-wal-12}, any element $\phi\in\mathbb{R}^3_\lambda$ has the following blockwise expansion%
\begin{equation}
\phi = \sum_{j\in\frac{\mathbb{N}}{2}} \ \sum_{-j\le m,n\in\mathbb{N}\le j} \phi^j_{mn} \ v^j_{mn} \ , \label{nat-fourier}
\end{equation}
where $\phi^j_{mn}\in\mathbb{C}$, and the family $\{v^j_{mn} \ , \ j\in\frac{\mathbb{N}}{2} \ ,\ -j\le m,n\le j\}$ is the natural orthogonal basis of $\mathbb{R}^3_\lambda$ introduced in \cite{vit-wal-12}, stemming from the direct sum decomposition%
\begin{equation}
\mathbb{R}^3_\lambda = \bigoplus_{j\in\frac{\mathbb{N}}{2}} \ \mathbb{M}_{2j+1}(\mathbb{C})\label{orthogonal-decomp}. 
\end{equation}
For fixed $j$, the corresponding subfamily is simply related to the canonical basis of the matrix algebra $\mathbb{M}_{2j+1}(\mathbb{C})$. The following fusion relation and conjugation hold true%
\begin{equation}
v^{j_1}_{mn} v^{j_2}_{qp} = \delta^{j_1j_2} \delta_{nq} \ v^{j_1}_{mp} \ , \ \ (v^j_{mn})^\dag=v^j_{nm} \ , \ \ 
\forall j\in\frac{\mathbb{N}}{2} \ , \ -j\le m,n,q,p\le j \ . \label{fusion}
\end{equation}
The orthogonality among the $v^j_{mn}$'s is taken with respect to the usual scalar product $\langle a,b\rangle:=\tr(a^\dag b)$, for any $a,b\in\mathbb{R}^3_\lambda$. Here, the trace functional $\tr$ can be defined \cite{gervitwal-13} for any $\Phi,\Psi\in\mathbb{R}^3_\lambda$ as%
\begin{equation}
\tr(\Phi\Psi) := 8 \pi \lambda^3 \sum_{j\in\frac{\mathbb{N}}{2}} w(j) \mbox{ tr}_j(\Phi^j\Psi^j)\label{trace-family}
\end{equation}
with $w(j)$ is a center-valued weight factor to be discussed below, $\mbox{ tr}_j$ denotes the canonical trace of $\mathbb{M}_{2j+1}(\mathbb{C})$, and $\Phi^j$ (resp. $\Psi^j$) an element of $\mathbb{M}_{2j+1}(\mathbb{C})$ is simply defined from the expansion \eqref{nat-fourier} of $\Phi$ by the $(2j+1)\times(2j+1)$ matrix $\Phi^j:=(\phi^j_{mn})_{-j\le m,n\le j}$ (resp. $\Psi^j:= (\psi^j_{qp})_{-j\le q,p\le j}$). Therefore we have%
\begin{equation}
\tr(\Phi\Psi)  = 8 \pi \lambda^3 \sum_{j\in\frac{\mathbb{N}}{2}} w(j) \left( \sum_{-j\le m,n\le j}\phi^j_{mn}\psi^j_{nm}\right), \label{traceb} 
\end{equation}
and%
\begin{equation}
\mbox{tr}_j(v^j_{mn}) = \delta_{mn} \ , \ \
\langle v^{j_1}_{mn} , v^{j_2}_{pq} \rangle = 8 \pi \lambda^3 \sum_{j_1\in\frac{\mathbb{N}}{2}} w(j_1) \ \delta^{j_1j_2} \delta_{mp} \delta_{nq} \ . \label{ortho-normaliz}
\end{equation}
Eqn. \eqref{trace-family} defines a family of traces depending on the weight factor $w(j)$. Recall that the particular choice%
\begin{equation}
w(j)=j+1\label{weight-gromov}
\end{equation}
leads to a trace that reproduces the expected behavior{\footnote{For instance, observe that one easily obtains from \eqref{traceb} the expected volume of a sphere of radius $\lambda N$ with $\Phi^j=\Psi^j=\bbone_j$ and summing up to $j=\frac{N}{2}$. Namely, one obtains $8\pi\lambda^3\overset{N}{\underset{k=0}{\sum}}\left(\frac{k}{2}\right)(k+1)\simeq\frac43\pi\left(\lambda N\right)^3$.}} for the usual integral on $\mathbb{R}^3$ once the (formal) commutative limit is applied \cite{gervitwal-13}. For a general discussion on this point based on a noncommutative generalization of the Kustaanheimo-Stiefel map \cite{ksmap}, see \cite{pv-ksmap}. \par%
We define $x_\pm := x_1\pm i x_2$. Other useful relations \cite{vit-wal-12} that will be needed for computations in the ensuing analysis are%
\begin{eqnarray}
x_+ \ v^j_{mn} = \lambda \ \mathcal{F}(j,m) \ v^j_{m+1, n}  
&& v^j_{mn} \ x_+ = \lambda \ \mathcal{F}(j,-n) \ v^j_{m, n -1} \nonumber \\
x_- \ v^j_{mn} = \lambda \ \mathcal{F}(j,-m) \ v^j_{m-1,n}  
&& v^j_{mn} \ x_- = \lambda \ \mathcal{F}(j,n) \ v^j_{m,n +1} \nonumber \\
x_3 \ v^j_{mn} = \lambda \ m \ v^j_{mn}
&& v^j_{mn} \ x_3 = \lambda \ n \ v^j_{mn} \nonumber \\
x_0 \ v^j_{mn} = \lambda \ j \ v^j_{mn}
&& v^j_{mn} \ x_0 = \lambda \ j \ v^j_{mn} \ , \label{x0-commut}
\end{eqnarray}
where%
\begin{equation}
\mathcal{F}(j,m):=\sqrt{(j+m+1)(j-m)} \ . \label{fjm}
\end{equation}


\subsection{\texorpdfstring{Differential calculus on $\mathbb{R}^3_\lambda$ and gauge theory models}{Family of gauge theories}} \label{subsection22}

The construction of noncommutative gauge models can be conveniently achieved by using the general framework of the noncommutative differential calculus based on the derivations of an algebra which has been introduced a long ago \cite{mdv88-99}. The general framework can actually be viewed as a noncommutative generalization the Koszul approach of differential geometry \cite{koszul}. Mathematical details and some related applications to NCFT can be found in \cite{cgmw-20}. \par 
In the present paper, we consider as in \cite{gervitwal-13} the differential calculus generated by the Lie algebra of real inner derivations of $\mathbb{R}^3_\lambda$%
\begin{equation}
\mathcal{G} := \left\{D_\mu:= Ad_{\theta_\mu}= i \left[\theta_\mu, \cdot\right]\right\} \ ,  \ \ \theta_\mu := \frac{x_\mu}{\lambda^2} \ , \ \ \forall \mu = 1,2,3 \ , \label{inv-form-conn}
\end{equation}
where the inner derivation $D_\mu$ satisfy the following commutation relation%
\begin{equation}
\left[D_\mu,D_\nu\right] = -\frac{1}{\lambda} \epsilon_{\mu\nu\rho} D_\rho,\ \forall \mu,\nu,\rho = 1,2,3 \ . \label{Der}
\end{equation}
Denoting, for any $n\in\mathbb{N}$, by $\Omega^n_\mathcal{G}$ the space of $n-(\mathcal{Z}(\mathbb{R}^3_\lambda))$-linear) antisymmetric maps $\omega:\mathcal{G}^n\to\mathbb{R}^3_\lambda$, the corresponding $\mathbb{N}$-graded differential algebra is $(\Omega_\mathcal{G}^\bullet = \oplus_{n\in\mathbb{N}} \Omega^n_\mathcal{G},\ d,\ \times)$, with nilpotent differential $d:\Omega^n_\mathcal{G}\to\Omega^{n+1}_\mathcal{G}$ and product $\times$ on $\Omega_\mathcal{G}^\bullet$ defined for any $\omega\in\Omega^p_\mathcal{G}$ and $\rho\in\Omega^q_\mathcal{G}$ by
\begin{eqnarray}
d\omega(X_1,...,X_{p+1})&=&\sum_{k=1}^{p+1}(-1)^{k+1}X_k\omega(X_1,...,\vee_k,...,X_{p+1})\nonumber\\
&+&\sum_{1\le k<l\le p+1}(-1)^{k+l}\omega([X_k,X_l],...,\vee_k,...,\vee_l,...,X_{p+1})\label{differential},
\end{eqnarray}
\begin{equation}
\omega\times\rho(X_1,...,X_{p+q})=\frac{1}{p!q!}\sum_{\sigma\in\mathfrak{S}_{p+q}}\vert\sigma\vert\omega(X_{\sigma(1),...,X_{\sigma(p)}})
\rho(X_{\sigma(p+1),...,X_{\sigma(p+q)}})\label{innerformproduct},
\end{equation}
where the $X_i$'s are elements of ${\cal{G}}$ and $\vert\sigma\vert$ is the signature of the permutation $\sigma\in\mathfrak{S}_{p+q}$.\par 

Let $\mathbb{M}$ denotes a right-module over $\mathbb{R}^3_\lambda$. Recall that a connection on $\mathbb{M}$ can be defined as a linear map $\nabla:\mathcal{G}\times\mathbb{M}\to\mathbb{M}$ with%
\begin{equation*}
\nabla_X(ma) = \nabla_X(m)a + mXa \ , \ \ \nabla_{zX}(a)=z\nabla_X(a) \ , \ \ \nabla_{X+Y}(a)=\nabla_X(a)+\nabla_Y(a) \ ,
\end{equation*}
for any $a\in\mathbb{R}^3_\lambda$, any $m\in\mathbb{M}$, $z\in \mathcal{Z}(\mathbb{R}^3_\lambda)$ and any $X,Y\in\mathcal{G}$.\par%
As we are interested by noncommutative versions of $U(1)$ gauge theories, we assume from now on $\mathbb{M} = \mathbb{C}\otimes\mathbb{R}^3_\lambda$ which can be viewed as a noncommutative analog of the complex line bundle relevant for abelian ($U(1)$) commutative gauge theories. We further restrict ourself to hermitean connections{\footnote{Given a hermitean structure, says $h:\mathbb{M}\times\mathbb{M}\to\mathbb{R}^3_\lambda$, $\nabla$ is hermitean if $Xh(m_1,m_2)=h(\nabla_X(m_1),m_2)+h(m_1,\nabla_X(m_2))$, for any $X\in{\cal{G}}$, $m_1,m_2\in\mathbb{M}$.}} for the canonical hermitean structure given by $h(a_1,a_2)=a_1^\dag a_2$, $a_1,a_2 \in \mathbb{R}^3_\lambda$. \\
A mere application of the above definition yields%
\begin{eqnarray}
\nabla_{D_\mu}(a) &:=& \nabla_\mu(a) = D_\mu a + A_\mu a \ , \nonumber \\
A_\mu &:=& \nabla_\mu(\bbone) \ , \quad \mbox{ with } \ A_\mu^\dag = - A_\mu \ , \label{connection}
\end{eqnarray}
for $a\in\mathbb{R}^3_\lambda$ and $\mu = 1,2,3$. The definition of the curvature 
\begin{equation*}
F(X,Y) := \left[\nabla_X,\nabla_Y\right] - \nabla_{\left[X,Y\right]} \ , \quad \forall X,Y \in \mathcal{G} \ , 
\end{equation*}
yields%
\begin{equation}
F(D_\mu,D_\nu) := F_{\mu\nu} = \left[\nabla_\mu,\nabla_\nu\right] - \nabla_{\left[D_\mu,D_\nu\right]} = D_\mu A_\nu - D_\nu A_\mu + \left[A_\mu,A_\nu\right] + \frac{1}{\lambda} \epsilon_{\mu\nu\rho} A_\rho \ . \label{curv1}
\end{equation}
The group of gauge transformations, defined as the group of automorphisms of the module compatible with both hermitean and right-module structures, is easily found to be the group of unitary elements of $\mathbb{R}^3_\lambda$, $\mathcal{U}(\mathbb{R}^3_\lambda)$, with left action of $\mathbb{R}^3_\lambda$. For any $g\in\mathcal{U}(\mathbb{R}^3_\lambda)$ and $\phi\in\mathbb{R}^3_\lambda$, one has $g^\dag g=gg^\dag=\bbone$, $\phi^g=g\phi$. From the definition of the gauge transformations of the connection given by $\nabla_\mu^g=g^\dag\nabla_\mu\circ g$, for any $g\in\mathcal{U}(\mathbb{R}^3_\lambda)$, one infers%
\begin{equation}
A_\mu^g = g^\dag A_\mu \ g + g^\dag D_\mu \ g \ , \ \ \mbox{and } \quad F^g_{\mu\nu} = g^\dag F_{\mu\nu} \ g \ \label{jauge-transfo}.
\end{equation}
The existence of a canonical gauge invariant connection, denoted hereafter by $\nabla^{inv}$, stems from the existence of inner derivations in the Lie algebra of derivations that generates the differential calculus. See \cite{mdv88-99} for a general analysis. In the present case, one finds
\begin{equation} 
\nabla^{inv}_\mu(a) = D_\mu a - i \theta_\mu a = - i a \theta_\mu \ , \quad \forall a \in \mathbb{R}^3_\lambda \ , \label{invar-connect}
\end{equation}
with curvature $F^{inv}_{\mu\nu}=0$. A natural gauge covariant tensor 1-form is then obtained by forming the difference between $\nabla^{inv}_\mu$ and any arbitrary connection. The corresponding components, sometimes called covariant coordinates, are given by%
\begin{equation}
\mathcal{A}_\mu := \nabla_\mu - \nabla^{inv}_\mu = A_\mu + i \theta_\mu \ , \quad \forall i=1,2,3 \ , \label{tens-form}
\end{equation}
and one has $\mathcal{A}_\mu^\dag = - \mathcal{A}_\mu$, $\mu=1,2,3$ ($A_\mu^\dag=-A_\mu$). By using \eqref{curv1}, one obtains%
\begin{equation} 
F_{\mu\nu} = \left[\mathcal{A}_\mu,\mathcal{A}_\nu\right] + \frac{1}{\lambda} \epsilon_{\mu\nu\rho} \mathcal{A}_\rho \ . \label{curv2}
\end{equation}
One easily verifies that for any $a\in\mathbb{R}^3_\lambda$, and $g\in\mathcal{U}(\mathbb{R}^3_\lambda)$, the following gauge transformations hold true%
\begin{equation}
(\nabla^{inv}_\mu(a))^g = \nabla^{inv}_\mu(a) \ , \quad \mathcal{A}^g_\mu = g^\dag \mathcal{A}_\mu \ g \ , \quad \forall \mu=1,2,3 \ . \label{conection-invariace}
\end{equation}
Define the real invariant 1-form $\Theta\in\Omega^1_\mathcal{G}$ by%
\begin{equation}
\Theta \in \Omega^1_\mathcal{G} \ : \ \Theta(D_\mu) = \Theta(Ad_{\theta_\mu}) = \theta_\mu \ .
\end{equation}
By making use of \eqref{differential} and \eqref{innerformproduct}, one easily check that%
\begin{equation}
d(-i\Theta)+(-i\Theta)^2=0 \ ,
\end{equation}
reflecting $F^{inv}_{\mu\nu}=0$. \\ 
The form $\Theta$ related to the 1-form invariant canonical connection supports an interesting interpretation. Recall \cite{mdv88-99} that a natural noncommutative analog of a symplectic form is defined as a real closed 2-form $\omega$ such that for any element $a$ in the algebra, there exists a derivation $\ham(a)$ (the analog of Hamiltonian vector field) verifying $\omega(X,\ham(a))=X(a)$ for any derivation $X$. One then observes that $\omega:=d\Theta\in\Omega^2_\mathcal{G}$ can be viewed as the natural symplectic form on the algebra $\mathbb{R}^3_\lambda$ in the setting of \cite{mdv88-99} with $\ham(a)=Ad_{ia}$ for any $a\in\mathbb{R}^3_\theta$ as the noncommutative analog of Hamiltonian vector field and 
\begin{equation}
\left\{a,b\right\}:=\omega\left(\ham(a),\ham(b)\right)=-i\left[a,b\right]\label{poissonB}
\end{equation}
the related (real) Poisson bracket.\par

\subsection{A family of gauge invariant classical actions}

Families of gauge-invariant functional (classical) actions can be easily obtained from the trace of any gauge-covariant polynomial functional in the covariant coordinates $\mathcal{A}_\mu$, namely $S_{inv}(\mathcal{A}_\mu)=\tr\left(P(\mathcal{A}_\mu)\right)$. Here, we will assume that the relevant field variable is $\mathcal{A}_\mu$, akin to a matrix model formulation of gauge theories on $\mathbb{R}^3_\lambda$, thus proceeding in the spirit of \cite{MVW13}. Natural requirement for the gauge-invariant functional are:
\begin{enumerate}[i)]
\item $P(\mathcal{A}_\mu)$ is at most quartic in $\mathcal{A}_\mu$,
\item $P(\mathcal{A}_\mu)$ does not involve linear term in $\mathcal{A}_\mu$ (not tadpole at the classical order),
\item the kinetic operator is positive.
\end{enumerate}
Set from now on 
\begin{equation*}
x^2 := \sum_{\mu=1}^3x_\mu x_\mu. 
\end{equation*}
We observe that gauge theories on $\mathbb{R}^3_\lambda$ can accommodate a 
gauge-invariant harmonic term $\sim\tr(x^2\mathcal{A}_\mu \mathcal{A}_\mu)$. This property simply stems from the fact that $x^2\in{\cal{Z}}(\mathbb{R}^3_\lambda)$ combined with the gauge-invariance of the 1-form canonical connection whose components in the module are given by 
\begin{equation}
\nabla^{inv}(\bbone)_\mu:=A^{inv}_\mu=-i\theta_\mu\label{component-invar-connect}
\end{equation}
as it can be readily obtained from \eqref{connection} and \eqref{invar-connect}. One easily checks that 
\begin{equation}
(A^{inv}_\mu)^g=(-i\theta_\mu)^g=-i\theta_\mu\label{invariance-component},
\end{equation}
as a mere combination of \eqref{inv-form-conn} and \eqref{jauge-transfo}. Now, the relation $\mathcal{R}_2$ \eqref{defining} and \eqref{inv-form-conn} imply
\begin{equation}
\sum_{\mu=1}^3(-i\theta_\mu)(-i\theta_\mu ) = -\frac{1}{\lambda^4}x^2 = -\frac{1}{\lambda^4}(x_0^2+\lambda x_0) \ , \label{harmonic-operator}
\end{equation}
in which the LHS is obviously gauge-invariant since \eqref{invariance-component} holds true while the RHS belongs to ${\cal{Z}}(\mathbb{R}^3_\lambda)$ as a polynomial in $x_0$. Hence, the gauge-invariant object $\sum_{\mu=1}^3(-i\theta_\mu)^2$ belongs to the center of $\mathbb{R}^3_\lambda$. Therefore, by using the cyclicity of the trace, one can write (summation over repeated $\alpha$ indice understood) %
\begin{eqnarray}
\tr(\sum_{\mu=1}^3(-i\theta_\mu)^g (-i\theta_\mu)^g (\mathcal{A}^g_\alpha \mathcal{A}^g_\alpha) ) &=& 
\tr( g \sum_{\mu=1}^3(-i\theta_\mu)(-i\theta_\mu) g^\dag (\mathcal{A}_\alpha \mathcal{A}_\alpha) )\nonumber\\
&=& \tr( \sum_{\mu=1}^3(-i\theta_\mu)(-i\theta_\mu) (\mathcal{A}_\alpha \mathcal{A}_\alpha) )
\label{harm-inv-dem}
\end{eqnarray}
where we used $\sum_\mu(-i\theta_\mu)(-i\theta_\mu)\in{\cal{Z}}(\mathbb{R}^3_\lambda)$ to obtain the last equality. Note that such a gauge-invariant harmonic term cannot be built in the case of gauge theories on the Moyal space $\mathbb{R}^4_\theta$ \cite{Wallet:2007c} simply because, says $x_{\nu=1,2,3,4}^2$, while still related to a gauge invariant object (a canonical gauge-invariant connection still exists, see e.g \cite{cgmw-20}), does not belong to the center of $\mathbb{R}^4_\theta$. \par%
It is convenient to work with hermitean fields. Thus, we set from now on
\begin{equation*}
\mathcal{A}_\mu = i \Phi_\mu
\end{equation*}
so that $\Phi^\dag_\mu = \Phi_\mu$ for any $\mu=1,2,3$. The above observation, combined with the requirements i) and ii) given above points towards the following general expression for a gauge-invariant action%
\begin{eqnarray}\label{fullaction}
S(\Phi)&=&\frac{1}{g^2} \tr\big( \kappa \Phi_\mu \Phi_\nu \Phi_\nu \Phi_\mu + \eta \Phi_\mu \Phi_\nu \Phi_\mu \Phi_\nu + i \zeta \epsilon_{\mu\nu\rho} \Phi_\mu \Phi_\nu \Phi_\rho + (M+\mu x^2) \Phi_\mu \Phi_\mu \big)\nonumber\\
&=&\frac{1}{g^2} \tr\big((\frac{\eta-\kappa}{4})[\Phi_\mu,\Phi_\nu]^2+(\frac{\eta+\kappa}{4})\{\Phi_\mu,\Phi_\nu \}^2
+ i \zeta \epsilon_{\mu\nu\rho} \Phi_\mu \Phi_\nu \Phi_\rho\nonumber\\
&+& (M+\mu x^2) \Phi_\mu \Phi_\mu \big), \label{class-action}
\end{eqnarray}
where from now on Einstein summation convention is used, the trace is still given by \eqref{traceb} and $g^2$, $\kappa$, $\eta$, $\zeta$, $M$ and $\mu$ are real parameters. The corresponding mass dimensions are
\begin{equation}
[\kappa]=[\eta]=0,\ [g^2]=[\zeta]=1,\ [M]=2,\ [\mu]=4\label{mass-dim}
\end{equation}
so that the action \eqref{class-action} is dimensionless, assuming that the ``engineering'' dimension $3$ of the noncommutative space is the relevant dimension.\par%
We will mainly focus on sub-families involving positive actions obtained from \eqref{class-action}. In order to make contact with some notations of ref. \cite{Wallet:2007c}, we set%
\begin{equation}
\kappa= 2(\Omega+1),\ \eta= 2(\Omega-1), \label{redef-param}
\end{equation}
where the real parameter $\Omega$ is dimensionless, thus fixing for convenience the overall normalization of the term $\sim[\Phi_\mu,\Phi_\nu]^2$ in \eqref{class-action}. This latter action can be rewritten as
\begin{eqnarray}
S(\Phi)&=&\frac{1}{g^2} \tr\big((F_{\mu\nu} - \frac{i}{\lambda} \epsilon_{\mu\nu\rho} \Phi_\rho)^\dag (F_{\mu\nu} - \frac{i}{\lambda} \epsilon_{\mu\nu\rho} \Phi_\rho) + \Omega\left\{\Phi_\mu,\Phi_\nu\right\}^2
+i \zeta \epsilon_{\mu\nu\rho} \Phi_\mu \Phi_\nu \Phi_\rho\nonumber\\
&+& (M+\mu x^2) \Phi_\mu \Phi_\mu \big)\nonumber\\
&=&\frac{1}{g^2} \tr\big(F^\dag_{\mu\nu}F_{\mu\nu} + \Omega\left\{\Phi_\mu,\Phi_\nu\right\}^2 + i \zeta^\prime\epsilon_{\mu\nu\rho} \Phi_\mu \Phi_\nu \Phi_\rho + \left(M^\prime+\mu x^2\right) \Phi_\mu \Phi_\mu \big)\label{canonic-action},
\end{eqnarray}
with
\begin{equation}
\zeta = \zeta^\prime+\frac{4}{\lambda};\ \ M=M^\prime+\frac{2}{\lambda^2}. \label{new-param}
\end{equation}
We note that the first two terms in the gauge-invariant action $S(\Phi)$ \eqref{canonic-action} are formally similar to those occurring in the so-called induced gauge theory on $\mathbb{R}^4_\theta$ \cite{Wallet:2007c}.\\
$S(\Phi)$ is positive when 
\begin{equation}
\Omega\ge0,\ \mu>0,\ \zeta=0,\ M>0\label{thecondition-positivity}
\end{equation}
or
\begin{equation}
\Omega\ge0,\ \mu>0,\ \zeta=\frac{4}{\lambda},\ M>\frac{2}{\lambda^2}, 
\end{equation}
as it can be realized respectively from the 1st and 2nd equality in \eqref{canonic-action} (see also section \ref{section3} and the appendix for the positivity of the kinetic operator).\par

In the rest of this paper, we will focus on the family of actions fulfilling the first condition \eqref{thecondition-positivity}, namely
\begin{equation}
S_\Omega = \frac{1}{g^2} \tr\big((F_{\mu\nu} - \frac{i}{\lambda} \epsilon_{\mu\nu\rho} \Phi_\rho)^\dag (F_{\mu\nu} - \frac{i}{\lambda} \epsilon_{\mu\nu\rho} \Phi_\rho) + \Omega\left\{\Phi_\mu,\Phi_\nu\right\}^2 + (M+\mu x^2) \Phi_\mu \Phi_\mu \big) \label{zeta=0}.
\end{equation}
The equation of motion for \eqref{zeta=0} given by
\begin{equation}
4(\Omega+1)(\Phi_\rho\Phi_\mu\Phi_\mu+\Phi_\mu\Phi_\mu\Phi_\rho)+8(\Omega-1)\Phi_\mu\Phi_\rho\Phi_\mu+2(M+\mu x^2) \Phi_\rho = 0\label{eqn-motion},
\end{equation}
one infers that $\Phi_\rho=0$ is the absolute minimum of \eqref{zeta=0}\footnote{There are also other nontrivial solutions of the equation of motion related to \eqref{fullaction}. Namely, there is one more solution belonging to the center ${\cal{Z}}(\mathbb{R}^3_\lambda)$ given by $\Phi_{\mu}\Phi_{\mu}=-\frac{M+\mu x^2}{2(\kappa+\eta)}$. We found also solution outside the center given by $\Phi_i=fx_i$, where $f=\frac{-\eta\lambda\pm\sqrt{\eta^2\lambda^2-32\left[x^2(\kappa+\eta)-\eta\lambda^2\right](M+\mu x^2)}}{8\left[x^2(\kappa+\eta)-\eta\lambda^2\right]}$. The corresponding quantum field theories are still under investigation. }.\par

In the section \ref{section3}, we will show that one class of gauge-invariant models pertaining to the families \eqref{zeta=0}, \eqref{canonic-action} yields after gauge-fixing to a finite theory at all orders in perturbation. This stems from the conjunction of the gauge-invariant harmonic term in \eqref{class-action} $\sim \mu x^2\Phi_\mu\Phi_\mu$, the orthogonal sum structure of $\mathbb{R}^3_\lambda$ \eqref{orthogonal-decomp} and the existence of a bound on the (absolute value of) the propagator for $\Phi_\mu$. This will be discussed at the end of the paper. Notice that in the Moyal case only the term $\sim M$ is allowed by gauge invariance.


\section{Perturbative analysis.}\label{section3}

\subsection{Gauge-fixing.}\label{subsection31}

We set%
\begin{equation}
\Phi_\mu = \sum_{j,m,n} (\phi_\mu)^j_{mn} v^j_{mn} \ , \ \ \forall \mu=1,2,3.
\end{equation}
The kinetic term of the classical action \eqref{zeta=0} $S_\Omega$ is given by%
\begin{eqnarray}
S_{Kin}(\Phi) &=& \frac{1}{g^2} \tr( \Phi_\mu (M+\mu x^2) \Phi_\mu)\label{skin} \\
&=& \frac{8\pi\lambda^3}{g^2} \sum_{j,m,n} w(j) (M+\lambda^2\mu j(j+1)) |(\phi_\mu)^j_{mn}|^2 \label{skin-explicit}
\end{eqnarray}
where $w(j)$ is the center-valued weight introduced in \eqref{traceb}) and we used \eqref{fusion}, \eqref{x0-commut}, \eqref{traceb} and 
\begin{equation}
x_0 = \lambda\sum_{j,m} \ j \ v^j_{mm},\ x^2 = \lambda^2\sum_{j,m} \ j(j+1) \ v^j_{mm},\label{relation-apendix1}
\end{equation}
stemming from \eqref{x0-commut} and \eqref{nat-fourier}. Recall that we have assumed that the condition \eqref{thecondition-positivity} holds true. We assume for the moment that $w(j)$ is a polynomial function of $j$, thus insuring a suitable decay of the related propagators at large indices. We will specialize to the cases $w(j)=1$ and $w(j)=j+1$ in a while.\\
Now, defining the kinetic operator by
\begin{equation*}
S_{Kin}(\Phi)=\sum_{j,m,n,k,l}(\phi_\mu)^{j_1}_{mn}G^{j_1j_2}_{mn;kl}(\phi_\mu)^{j_2}_{kl},
\end{equation*}
one can write 
\begin{equation}
G^{j_1j_2}_{mn;kl} = \frac{8\pi\lambda^3}{g^2} w(j_1) \ \left(M+\lambda^2\mu j_1(j_1+1)\right) \delta^{j_1j_2} \delta_{nk} \delta_{ml}.\label{kin-op1}
\end{equation}
The relation \eqref{kin-op1} defines a positive self-adjoint operator. The corresponding details are collected in the appendix \ref{operators}.\par 

The gauge-invariance of $S_\Omega$ \eqref{zeta=0} can be translated into invariance under a nilpotent BRST operation $\delta_0$ defined by the following structure equations \cite{MVW13}%
\begin{equation}
\delta_0 \Phi_\mu = i [C,\Phi_\mu] \ , \ \ \delta_0C=iCC\label{brs}
\end{equation}
where $C$ is the ghost field. Recall that $\delta_0$ acts as an antiderivation with respect to the grading given by (the sum of) the ghost number (and degree of forms), modulo 2. $C$ (resp. $\Phi_i$) has ghost number $+1$ (resp. $0$). Fixing the gauge symmetry can be conveniently done by using the gauge condition 
\begin{equation}
\Phi_3=\theta_3\label{special-gauge}.
\end{equation}
This can be implemented into the action by enlarging \eqref{brs} with%
\begin{equation}
\delta_0 {\bar{C}} = b \ , \ \ \delta_0b = 0 \label{contractible-brs}
\end{equation}
where ${\bar{C}}$ and $b$ are respectively the antighost and the St\"uckelberg field (with respective ghost number $-1$ and $0$) and by adding to $S_\Omega$ a BRST invariant gauge-fixing term given by \eqref{zeta=0}%
\begin{equation}
S_{fix}=\delta_0\tr\big({\bar{C}}(\Phi_3-\theta_3) \big)=\tr\big(b(\Phi_3-\theta_3)-i{\bar{C}}[C,\Phi_3]\big)\label{gauge-fix}.
\end{equation}
Integrating over the St\"ueckelberg field $b$ yields the constraint $\Phi_3=\theta_3$ into \eqref{zeta=0}, while the ghost part can be easily seen to decouple{\footnote{Recall it amounts to consider an "on-shell" formulation for which nilpotency of the BRST operation (and corresponding BRST-invariance of the gauge-fixed action) is verified modulo the ghost equation of motion.}}. \par 
Now, we define the kinetic operator by%
\begin{equation}
K:=G+8\Omega L(\theta_3^2)\label{operator-K}.
\end{equation}
where $G=M+\mu x^2$ and $L(\theta^{2}_{3})$ is the left multiplication by $\theta^{2}_{3}$.
The resulting gauge-fixed action can be written (up to an unessential constant term) as
\begin{equation}
S^f_\Omega=S_2+S_4\label{stot},
\end{equation}
with%
\begin{eqnarray}
S_2 &=& \frac{1}{g^2} \tr ((\Phi_1,\Phi_2)
\begin{pmatrix}
Q&0\\
0&Q
\end{pmatrix} 
\begin{pmatrix}
\Phi_1\\
\Phi_2
\end{pmatrix} 
) , \nonumber \\
Q &=& K + i4 (\Omega-1) L(\theta_3) D_3 \ , \label{squad1} \\[5pt]
S_4 &=& \frac{4}{g^2} \tr \left( \Omega (\Phi_1^2 + \Phi_2^2)^2 + (\Omega-1)(\Phi_1\Phi_2\Phi_1\Phi_2 - \Phi_1^2\Phi_2^2) \right) . \label{squart}
\end{eqnarray}
The gauge-fixed action \eqref{stot} is thus described by a rather simple NCFT with "flavor diagonal" kinetic term (see \eqref{squad1}) and quartic interaction terms. We find also convenient to introduce the complex fields%
\begin{equation}
\Phi=\frac{1}{2}(\Phi_1+i\Phi_2),\ \Phi^\dag=\frac{1}{2}(\Phi_1-i\Phi_2),
\end{equation}
so that the gauge-fixed action $S^f_\Omega$ can be expressed alternatively into the form%
\begin{equation}
S^f_\Omega = \frac{2}{g^2} \tr\left( \Phi Q \Phi^\dag + \Phi^\dag Q\Phi \right) + \frac{16}{g^2} \tr\left( (\Omega+1) \Phi\Phi^\dag\Phi\Phi^\dag + (3\Omega-1) \Phi\Phi\Phi^\dag\Phi^\dag \right) .
\label{quasilsz}
\end{equation}
At this level, some comments are in order.
\begin{itemize}
\item The action \eqref{quasilsz} bears some similarity with the (matrix model representation of) the action describing the family of complex LSZ models \cite{LSZ}.%

\item For $\Omega=1/3$, the quartic interaction potential depends only on $\Phi\Phi^\dag$, so that the action is formally similar to the action describing an exactly solvable LSZ-type model investigated in \cite{LSZ}. Only the respective kinetic operators are different. It turns out that the partition function for $S^f_{\Omega=\frac{1}{3}}$ \eqref{quasilsz} can be actually related to $\tau$-functions of integrable hierarchies. More precisely, due to the orthogonal decomposition of $\mathbb{R}^3_\lambda$ 
\eqref{orthogonal-decomp}, the partition function can be expressed as a product of factors labelled by $j\in\frac{\mathbb{N}}{2}$, each one being expressible as a $\tau$-function for a 2-d Toda hierarchy. Note that each factor can be actually interpreted as the partition function for the reduction of the gauge-fixed theory \eqref{stot} on the matrix algebra $\mathbb{M}_{2j+1}(\mathbb{C})$. The corresponding analysis will be presented in a separate publication \cite{solvab-15}.

\item For $\Omega=1$, the kinetic operator in \eqref{quasilsz} simplifies while the interaction term takes a more symmetric form, as it is apparent e.g from \eqref{squart}. We will find that the corresponding theory is finite to all orders in perturbation.%
\end{itemize}

\subsection{\texorpdfstring{Gauge-fixed action at $\Omega=1$.}{The Omega=1 case}}\label{subsection32}

In this subsection, we will assume $\Omega=1$. The corresponding action is%
\begin{equation}
S^f_{\Omega=1} = \frac{1}{g^2} \tr( (\Phi_1,\Phi_2)
\begin{pmatrix}
K&0\\
0&K
\end{pmatrix} 
\begin{pmatrix}
\Phi_1\\
\Phi_2
\end{pmatrix} 
)
+ \frac{4}{g^2} \tr( (\Phi_1^2 + \Phi_2^2)^2).\label{critical-action}
\end{equation}
The kinetic term is expressed as 
\begin{equation}
S^f_{2, \Omega=1}=\frac{8\pi\lambda^3}{g^2}\sum_{j,m,n}w(j)(M+\mu\lambda^2j(j+1)+\frac{8}{\lambda^2}n^2)\vert(\phi_{1\mu})_{mn}\vert^2+(1\to2)\label{kin-omega1},
\end{equation}
where we used
\begin{equation}
x_3^2 = \lambda^2 \sum_{j,m} m^2 v^j_{mm}.
\end{equation}
Accordingly, the "matrix elements" of the kinetic operator can be written as
\begin{equation}
K^{j_1 j_2}_{mn;kl} := \frac{8\pi\lambda^3}{g^2} w(j_1) ( M + \mu \lambda^2 j_1 (j_1+1) + \frac{4}{\lambda^2} (k^2+l^2) ) \delta^{j_1j_2} \delta_{ml} \delta_{nk}. \label{matrix-K} 
\end{equation}
Note that \eqref{matrix-K} verifies%
\begin{equation}
K^{j_1j_2}_{mn;kl} = K^{j_1j_2}_{lk;nm} = K^{j_1j_2}_{mn;lk} \label{sym-K}
\end{equation}
reflecting reality of the functional action and the self-adjointness of $K$ (see appendix \ref{operators}; recall we use the natural Hilbert product $\langle a,b \rangle = \tr(a^\dag b)$).\par

The inverse of \eqref{matrix-K} (i.e the matrix elements of the propagator) $P^{j_1j_2}_{mn;kl}$ is then defined by%
\begin{equation}
\sum_{j_2,k,l} K^{j_1j_2}_{mn;lk} P^{j_2j_3}_{kl;rs} = \delta^{j_1j_3} \delta_{ms} \delta_{nr}, \ \ \sum_{j_2,n,m} P^{j_1j_2}_{rs;mn} K^{j_2j_3}_{nm;kl} = \delta_{j_1j_3} \delta_{rl} \delta_{sk}, \label{propagator-def}
\end{equation}
leading to%
\begin{equation}
P^{j_1j_2}_{mn;kl} = \frac{g^2}{8\pi\lambda^3} \frac{1}{w(j_1)(M+\lambda^2\mu j_1(j_1+1)+\frac{4}{\lambda^2}(k^2+l^2))}\delta^{j_1j_2}\delta_{ml}\delta_{nk}.  \label{propagator}
\end{equation}

We will start the perturbative analysis by computing the 2-point (connected) correlation function at the first (one-loop) order. To prepare the discussion, we introduce sources variables for the $\Phi_\alpha$'s, namely $J_\alpha = \underset{j,m,n}{\sum}(J_\alpha)^j_{mn}v^j_{mn}$, for any $\alpha=1,2$. Then, a standard computation yields the free part of the generating functional of the connected correlation functions $W_0(J)$ given (up to an unessential prefactor) by%
\begin{eqnarray}
e^{W_0(J)}&=& \int\prod_{\alpha=1}^2\mathcal{D}\Phi_\alpha e^{-(S^f_{2\Omega=1}+\tr(\Phi_\alpha J_\alpha))}
=\int\prod_{\alpha=1}^2\mathcal{D}\Phi_\alpha e^{-\sum((\phi_\alpha)^{j_1}_{mn}K^{j_1j_2}_{mn;kl}(\phi_\alpha)^{j_2}_{kl}+(\mathcal{J}_\alpha)^j_{mn}(\phi_\alpha)^{j}_{nm})}\nonumber\\
&=&\exp(\frac{1}{4}\sum(\mathcal{J}_\alpha)^{j_1}_{mn}P^{j_1j_2}_{mn;kl}(\mathcal{J}_\alpha)^{j_2}_{kl})\label{free-generat},
\end{eqnarray}
where we have defined for further convenience
\begin{equation}
(\mathcal{J}_\alpha)^{j}:=8\pi\lambda^3w(j)({J}_\alpha)^{j}, -j\le m,n\le j\label{source}
\end{equation}
for any $j\in\frac{\mathbb{N}}{2}$. To obtain \eqref{free-generat}, one simply uses the generic field redefinition among the fields components given by%
\begin{equation*}
(\phi_\alpha)^j_{mn} \ = \ (\phi^\prime_\alpha)_{mn}^j - \frac12 {P}^j_{nm;kl} (\mathcal{J}_\alpha)^j_{kl} \ = \ (\phi^\prime_\alpha)_{mn}^j - \frac12 (\mathcal{J}_\alpha)^j_{rs}{P}_{rs;nm}. 
\end{equation*}
Correlation functions involving modes $(\phi_\alpha)^j_{mn}$ will be obtained from the successive action of the corresponding functional derivatives $\frac{\delta}{\delta(\mathcal{J}_\alpha)^j_{nm}}$ on the full generating functional. We use
\begin{equation}
e^{-S_4(\Phi_1,\Phi_2)}e^{-\tr(J_\alpha\Phi_\alpha)}=e^{-S_4(\frac{\delta}{\delta\mathcal{J}_1},\frac{\delta}{\delta\mathcal{J}_2})}e^{-\sum(\mathcal{J}_\alpha)^j_{mn}(\phi_\alpha)^{j}_{nm}}\label{trick}
\end{equation}
where
\begin{equation}
S_4(\frac{\delta}{\delta\mathcal{J}_1},\frac{\delta}{\delta\mathcal{J}_2})=\sum\frac{8\pi\lambda^3}{g^2}w(j)S^j_4(\frac{\delta}{\delta\mathcal{J}})\label{s4intfunction}
\end{equation}
in which $S^j_4$ denotes the $\tr_j$ part of the interaction term in the action \eqref{critical-action}. We then write 
\begin{equation*}
e^{W(\mathcal{J})} = e^{-S_4(\frac{\delta}{\delta\mathcal{J}_1},\frac{\delta}{\delta\mathcal{J}_2})}\ e^{W_0(\mathcal{J})}
\end{equation*}
to obtain%
\begin{equation}
W(\mathcal{J}) = W_0(\mathcal{J}) + \ln\left[ 1 + e^{-W_0(\mathcal{J})}
\left( e^{-S_4(\frac{\delta}{\delta\mathcal{J}_1},\frac{\delta}{\delta\mathcal{J}_2})} - 1 \right) e^{W_0(\mathcal{J})}\right] \ , \label{connected-funct}
\end{equation}
where $S_4$ is defined by \eqref{s4intfunction}. The expansion of both the logarithm and $e^{S_4}$ then gives rise to the perturbative expansion. \par 

\subsection{One-loop 2-point and 4-point functions.}\label{subsection322}

The computational details of the one-loop contribution to the 2-point function are collected in the appendix \ref{2-point-comput}. From \eqref{B-Gamma}, it can be realized that the quadratic part of the classical action receives a 1st order (one-loop) contribution $\Gamma^1_2(\Phi_\alpha)$ given by
\begin{eqnarray}
\Gamma^1_2(\Phi_\alpha)=\frac{32\pi\lambda^3}{g^2}\sum_{j\in\frac{\mathbb{N}}{2}}&\Bigg[&\sum_{-j\le m,n,r,p\le j}(\phi_\alpha)^j_{pr}\left(w(j)P^j_{rm;np}\right)(\phi_\alpha)^j_{mn}\nonumber\\
&+&\sum_{-j\le p,r,n\le j}3(\phi_\alpha)^j_{pr}\left(\sum_{m=-j}^jw(j)P^j_{rm;mn}\right)(\phi_\alpha)^j_{np} \Bigg]\label{Gamma-oneloop},
\end{eqnarray}
in which the 1st (resp. 2nd) term corresponds to the non-planar (resp. planar) contribution. Writing generically $\Gamma^1_2(\Phi_\alpha)=\frac{32\pi\lambda^3}{g^2}\sum(\phi_\alpha)^j_{mn}\sigma^j_{mn;kl}(\phi_\alpha)^j_{kl}$, we have explicitly
\begin{eqnarray}
\sigma^{NP\ j}_{pr;mn}&=& w(j)P^j_{pr;mn}\label{eff-act-np} \\
\sigma^{P\ j}_{pr;nm}&=&3\delta_{mp}\sum_{m=-j}^jw(j)P^j_{rm;mn}\label{eff-act-plan}.
\end{eqnarray}

One can easily verify that \eqref{eff-act-plan} and \eqref{eff-act-np} are always finite, even for $j=0$ and $j\to\infty$ and without any singularity whenever $M>0$, which is assumed here. This is obvious for \eqref{eff-act-np}. For the planar contribution, one simply observes that the summation over $m$, which corresponds to an internal ribbon loop, satisfies the estimate%
\begin{eqnarray}
\sum_{m=-j}^jw(j)P_{rm;mn}&=&\delta_{nr}\sum_{m=-j}^j\frac{g^2}{8\pi\lambda^3}\frac{1}{(M+\lambda^2\mu j(j+1)+\frac{4}{\lambda^2}(m^2+n^2))}\nonumber\\
&\le& \delta_{nr}\frac{g^2}{8\pi\lambda^3}\frac{2j+1}{(M+\lambda^2\mu j(j+1))} \label{bound-1loop}
\end{eqnarray}
which is always finite for any $j\in\frac{\mathbb{N}}{2}$. Note that no dangerous UV/IR mixing shows up in the computation of the one-loop 2-point function.\par %

Eqn.\eqref{bound-1loop} reflects simply the existence of an estimate obeyed by the propagator \eqref{propagator} (see \eqref{envelop-model} below). This can be used in the subsection \ref{subsection33} to show the finitude of the theory to all orders in perturbation. Indeed, we have from \eqref{propagator}:
\begin{equation}
0\le P^{j_1j_2}_{mn;kl}\le \frac{\Pi(M,j_1)}{w(j_1)}\delta_{j_1j_2}\delta_{ml}\delta_{nk}, \label{envelop-model}
\end{equation}
for any $j_1,j_2\in\frac{\mathbb{N}}{2},\ -j_1\le m,n,k,l\le j_1,$, where
\begin{equation}
\Pi(M,j) := \frac{g^2}{8\pi\lambda^3}\frac{1}{(M+\lambda^2\mu j(j+1))}.\label{Pi}
\end{equation}

A similar analysis can be carried out for the 1-loop contributions to the 4-point function showing that those contributions are again finite. For instance, consider the vertex functional for one specie $\Phi_\alpha$, written generically as (no sum over $\alpha$)
\begin{equation}
\Gamma^1_4(\Phi_\alpha)=\sum_{m_i,n_i,r_i,s_i} V_{m_1,m_2,n_1,n_2,r_1,r_2,s_1,s_2}(\phi_\alpha)^j_{m_1m_2}(\phi_\alpha)^j_{n_1n_2}(\phi_\alpha)^j_{r_1r_2}(\phi_\alpha)^j_{s_1s_2}\label{vertex-functional}.
\end{equation}
Typical planar contributions to the vertex functional are of the form
\begin{eqnarray}
\Gamma^{P\ 1}_4&\sim&\sum\big(\sum_{-j\le p,q\le j}w^2(j)P^j_{n_1p;qr_2}P^j_{pm_2;s_1q}\delta_{m_1n_2}\big)\nonumber\\
&&\times\delta_{s_2r_1} (\phi_\alpha)^j_{m_1m_2}(\phi_\alpha)^j_{n_1n_2}(\phi_\alpha)^j_{r_1r_2}(\phi_\alpha)^j_{s_1s_2}\label{planar-4pt},
\end{eqnarray}
where the factor $w^2(j)$ comes from the 2 vertex contributions to the loop. One can easily check that
\begin{equation}
\sum_{-j\le p,q\le j}w^2(j)P^j_{n_1p;qr_2}P^j_{pm_2;s_1q}\le\delta_{n_1r_2}
\delta_{s_1m_2}(2j+1)\Pi(M,j)^2\label{bound-planar-4pt},
\end{equation}
which is finite for any value of $j$ and decays to $0$ as $j^{-3}$ when $j\to\infty$. \\
Other planar 1-loop contributions to the vertex function can be checked to be finite by using a similar argument.\\
There are 3 species of non-planar contributions with typical respective contributions being of the form
\begin{eqnarray}
\Gamma^1_{14}\sim\sum &\big(&w^2(j)P^j_{m_1n_2;s_1r_2}P^j_{n_1m_2;r_1s_2}\big)(\phi_\alpha)^j_{m_1m_2}(\phi_\alpha)^j_{n_1n_2}(\phi_\alpha)^j_{r_1r_2}(\phi_\alpha)^j_{s_1s_2}\label{gamma14},\\
\Gamma^1_{24}\sim\sum&\big(&\sum_pw^2(j)P^j_{m_1p;s_1r_2}P^j_{pn_2;r_1s_2}\delta_{m_2n_1}\big)\nonumber\\
&&\times(\phi_\alpha)^j_{m_1m_2}(\phi_\alpha)^j_{n_1n_2}(\phi_\alpha)^j_{r_1r_2}(\phi_\alpha)^j_{s_1s_2}\label{gamma24},\\
\Gamma^1_{34}\sim\sum&\big(&\sum_{p,q}w^2(j)P^j_{pm_2;qs_2}P^j_{n_1p;s_1q}
\delta_{m_1n_2}\delta_{s_2r_1} \big)\nonumber\\
&&\times(\phi_\alpha)^j_{m_1m_2}(\phi_\alpha)^j_{n_1n_2}(\phi_\alpha)^j_{r_1r_2}(\phi_\alpha)^j_{s_1s_2}\label{gamma34},
\end{eqnarray}
where obvious summations are not explicitly written. By further performing the summations over $p$ and $q$ in \eqref{gamma24}-\eqref{gamma34} thanks to the delta functions in the propagators $P^j_{mn;kl}$ \eqref{propagator}, we arrive easily at the following estimates:
\begin{eqnarray}
w^2(j)P^j_{m_1n_2;s_1r_2}P^j_{n_1m_2;r_1s_2}&\le&\Pi(M,j)^2\delta_{m_1r_2}\delta_{n_2s_1}\delta_{n_1s_2}\delta_{m_2r_1}\label{decadix1}\\
\sum_pw^2(j)P^j_{m_1p;s_1r_2}P^j_{pn_2;r_1s_2}&\le&\Pi(M,j)^2\delta_{m_1r_2}\delta_{r_1n_2}\label{decadix2}\\
\sum_{p,q}w^2(j)P^j_{pm_2;qs_2}P^j_{n_1p;s_1q}&\le&\Pi(M,j)^2\delta_{s_1s_2}\delta_{m_2n_1}\label{decadix3},
\end{eqnarray}
leading to finite non-planar contributions to the vertex functional \eqref{vertex-functional}. A similar conclusion holds true for the other non-planar contribution. Notice, by the way that the RHS of each of the relations \eqref{bound-planar-4pt} and \eqref{decadix1}-\eqref{decadix3} decay to zero as $j^{-4}$ for $j\to\infty$.\par

As for the 2-point function, the diagram amplitudes for the 4-point function are finite, thanks to the existence of the bound for the propagator \eqref{envelop-model} together with the fact that loop summation indices are bounded by $\pm j$. Summarizing the above 1-loop analysis, a simple inspection shows that no singularity can occur for $j=0$ within the present model (recall $M>0$) while the only source for divergence might come from the limit $j\to\infty$. But such divergences are prevented to occur thanks to the upper bound \eqref{envelop-model} and the decay of $\Pi(M,j)$ \eqref{Pi} at large $j$, namely $\Pi(M,j)\sim j^{-2}$ for $j\to\infty$ so that the model \eqref{critical-action} is finite at the one-loop order. In the next subsection, we will show that this property extends to any order of perturbation.

\subsection{Finitude of the diagram amplitudes to all orders.}\label{subsection33}

We first observe that \eqref{Pi} is related obviously to the propagator for the "truncated" gauge model obtained by simply dropping the field $\Phi_3$ in the action \eqref{zeta=0}. Notice that this latter {\it{formally}} may be viewed as resulting from the gauge choice $\Phi_3=0$ in \eqref{gauge-fix} instead of $\Phi_3=\theta_3$. For convenience, we quote here the expression for the propagator of the truncated theory which can be simply read off from the RHS of \eqref{envelop-model} and \eqref{Pi}:
\begin{equation}
(G^{-1})^{j_1j_2}_{mn;kl}=\delta^{j_1j_2}\delta_{mn}\delta_{kl}\frac{\Pi(M,j_1)}{w(j_1)}\label{g-1}
\end{equation}
which depends only on a single $j\in\frac{\mathbb{N}}{2}$, says $j_1$.\\

The "truncated model" belongs to one particular class of NCFT on $\mathbb{R}^3_\lambda$ among those which have been investigated in \cite{vit-wal-12} where it was shown that the models in this class are finite to all orders in perturbation. We first discuss useful property of this model.\\
The key observation is that the amplitude of any ribbon diagram depends only on one $j\in\frac{\mathbb{N}}{2}$. Indeed, observe e.g the $\delta^{j_1j_2}$ in the propagator \eqref{g-1} plus its $j$-dependence and the delta functions in any quartic vertex. These $\delta^{j_mj_k}$'s all boil down to a single one in the computation of any amplitude.\\

Since the propagator \eqref{g-1} depends on the bounded indices $m,n,...$ only through Kronecker delta's, the summations over the indices of any loop can be exactly carried out so that any ribbon loop contributes to a factor 
\begin{equation}
(2j+1)^\varepsilon,\ \varepsilon\le2 \label{exponent-epsilon}
\end{equation}
to a given amplitude. This can be understood from a simple inspection of the Kronecker delta's and the summations over the indices for a ribbon loop built from any $N$-point sub-diagram ${\cal{A}}_{m_1,n_1,...,m_N,n_N}$ and a propagator \eqref{g-1} that can be taken to be $(Q^{-1})^j_{m_1n_1;m_2n_2}$ without loss of generality. Namely, one has
\begin{equation}
\mathbb{A}_{m_3,n_3,...,m_N,n_N}=\sum_{-j\le m_1,n_1,m_2,n_2\le j}{\cal{A}}_{m_1,n_1,...,m_N,n_N}(Q^{-1})^j_{m_1n_1;m_2n_2}\label{loop-sums}.
\end{equation}
There are 4 summed (internal) indices related to the product of N delta's 
coming from the $N$-point sub-diagram by the 2 delta's of the propagator depending only on internal indices. Two summations 
can be trivially performed leading to N remaining delta functions. There are a priori 3 possibilities depending how the 2 remaining summed indices are distributed among the delta's: either a single delta depends only on one internal index, or one get a product of two such deltas, one of each internal index, or
the 2 summations combine 2 deltas among the $N$ one leading to $N-2$ remaining deltas. The details are given in the appendix \ref{appendix3}. Notice that the value $\varepsilon=2$ is obtained from purely algebraic and combinatorial arguments and represents actually the maximal power of the factor $2j+1$ any loop can contribute. A refinement of this analysis by taking into account indices conservation may well lower the maximal value of this exponent by one unit. Nevertheless, it turns out that the use of this somewhat crude maximal value in the ensuing analysis is sufficient to prove the finitude of arbitrary amplitudes. Summarizing the above discussion, it appears that the loop summations decouple from the related propagators in the computation of diagram amplitudes for the truncated model, so that any loop simply contribute by a power of $(2j+1)$ given by \eqref{exponent-epsilon}. This leads to a major simplification in the analysis of amplitudes of arbitrary order, as it will be shown in a while.\par

To end up with perturbative considerations within the truncated model, consider now a general ribbon diagram ${\cal{D}}$ related to this model{\footnote{Recall that any ribbon in such a diagram is made of two lines each carrying 2 bounded indices, says $m,n\in\{-j,...,j \}$. Thus, a ribbon carries 4 bounded indices (as the propagator \eqref{g-1}). Notice that there is a conservation of the indices along each line, as it can be seen by observing the delta function in the expression of the propagator \eqref{g-1}, each delta defining the indices affected to one line. For more details, see \cite{vit-wal-12}.}}. Any ribbon diagram built from the quartic vertices is characterized by a set of positive integer $(V,I,F,B)$. $V$ is the number of vertices, $I$ the number of internal ribbons. $F$ is the number of faces. Recall that $F$ is obtained by closing the external lines of a diagram and counting the number of closed {\it{single}} lines. Finally, $B$ is the number of boundaries which is equal to the number of closed lines with external legs. The number of ribbon loops if given by
\begin{equation}
\mathcal{L}=F-B \label{loops-number}.
\end{equation}
Let $g\in\mathbb{N}$ be the genus of the Riemann surface on which $\mathcal{D}$ can be drawn. Recall that $g$ is determined by the following relation
\begin{equation}
2-2g=V-I+F \label{euler}.
\end{equation}
Now consider the amplitude ${\mathbb{A}}^{\cal{D}}$ for a diagram characterized by the parameters $(V,I,F,B)$. It is a (positive) function of $j$, obviously finite and non singular for $j=0$, built from the product of $V$ vertex factors, each vertex contributing to $w(j)$ up to unessential finite factor, $I$ propagators \eqref{g-1} with summations over indices corresponding to $F-B$ loops which, by the decoupling argument discussed above, give a net overall factor bounded by $(2j+1)^{2(F-B)}$. Therefore, we can write
\begin{equation}
{\mathbb{A}}^{\cal{D}}\le Kw(j)^{V-I}\Pi(M,j)^{I}(2j+1)^{2(F-B)}=K^\prime\frac{w(j)^{V-I}(2j+1)^{2(F-B)}}{(j^2+\rho^2)^I}\label{estim-amplt}
\end{equation}
where $K$ and $K^\prime$ are finite constants and $\rho^2=\frac{M}{\lambda\mu^2}$ and we have isolated the factor $w(j)$. Recall that the choice $w(j)=j+1$ as given in \eqref{weight-gromov} leads to a trace reproducing at the formal commutative limit the expected behavior for the usual integral on $\mathbb{R}^3$. The natural choice $w(j)=1$ is related to a functional trace built from all the canonical traces of the components $\mathbb{M}_{2j+1}(\mathbb{C})$ occurring in the decomposition of $\mathbb{R}^3_\lambda$, \eqref{orthogonal-decomp}. To study both cases when taking the $j\to\infty$ of the RHS of \eqref{estim-amplt}, we will set conveniently
\begin{equation}
w(j)\sim j^\alpha,\ \alpha=0,1,\ {\text{for}}\  j\to\infty.\label{alpha-values}
\end{equation}
The RHS of \eqref{estim-amplt} is always finite for $j=0$ while it is also finite for $j\to\infty$ provided%
\begin{equation}
\omega(\mathcal{D})=\alpha I+2B+2(2g-2)+V(2-\alpha)\ge0,\label{power-count}
\end{equation}
where we used \eqref{euler} and one has still $\alpha=0,1$. For $g\ge1$, one has $\omega(\mathcal{D})>0$. The case $g=0$, for which the finitude condition \eqref{power-count} becomes $\omega(\mathcal{D})=\alpha I+2B+V(2-\alpha)-4\ge0$ requires a closer analysis. In fact, when $V=2$ a simple inspection shows that \eqref{power-count} holds true for $\alpha=0,1$. The case $V=1$ corresponds to the 2-point function for the truncated model whose finitude when $j\to\infty$ is almost apparent from the rightmost quantity in \eqref{bound-1loop}. Note that this can be obtained from simple topological consideration for the planar and non planar contributions to this 2-point function. One obtains $B=2$ and $B=1$ respectively so that \eqref{power-count} holds true whenever $V=1$ for $\alpha=0,1$. Summarizing the above analysis, we conclude that the truncated model in finite to all orders in perturbation.\par

Let us go back to the gauge model \eqref{critical-action}. As far as finitude of the diagrams is concerned{\footnote{We consider only the finitude of the loop contributions and not the nature of the various vertices generated by loop corrections (i.e external legs) which simply amounts to analyze planar and non-planar contribution for a $\phi^4$ theory either with propagator \eqref{propagator} or with \eqref{g-1}}} one observes that \eqref{critical-action} differs from the truncated model only through the propagator. Hence, for a given diagram ${\cal{D}}$, the amplitude computed within the gauge model \eqref{critical-action} $\mathfrak{A}^j_{\mathcal{D}}$ satisfies 
\begin{equation}
\vert \mathfrak{A}^j_{\mathcal{D}}\vert\le \vert \mathbb{A}^j_{\mathcal{D}}\vert,
\end{equation}
thanks to the estimate \eqref{envelop-model}. Indeed, by using the general expression for any ribbon amplitudes of NC $\phi^4$ theory, one infers $\mathfrak{A}^j_{\mathcal{D}}$ has the generic structure%
\begin{equation}
\mathfrak{A}^j_{\mathcal{D}} = \sum_{\mathcal{I}} \prod_\lambda P^j_{m_\lambda(\mathcal{I}) n_\lambda(\mathcal{I});k_\lambda(\mathcal{I}) l_\lambda(\mathcal{I})} F^j(\delta)_{m_\lambda(\mathcal{I}) n_\lambda(\mathcal{I});k_\lambda(\mathcal{I}) l_\lambda(\mathcal{I})},\label{amplit-arb}
\end{equation}
where $\mathcal{I}$ is some set of (internal) indices, all belonging to $\{-j,...j\}$ so that all the sums in $\sum_{{\cal{I}}}$ are finite, $\lambda$ labels the internal lines of ${\cal{D}}$, $P^j_{mn;kl}$ is the (positive) propagator given in \eqref{propagator} and $F^j(\delta)_{mn;kl}$ collects all the delta's plus vertex weights depending only on $j$. One has%
\begin{eqnarray}
\vert\mathfrak{A}^j_{\mathcal{D}}\vert &\le& \sum_{\mathcal{I}} \prod_\lambda \left|(G^{-1})^j_{m_\lambda(\mathcal{I}) n_\lambda(\mathcal{I}) ; k_\lambda(\mathcal{I}) l_\lambda(\mathcal{I})} \right| \ \left| F^j(\delta)_{m_\lambda({\cal{I}}) n_\lambda({\cal{I}});k_\lambda({\cal{I}}) l_\lambda({\cal{I}})} \right|.
\end{eqnarray}
From \eqref{power-count}, one then obtains
\begin{equation}
\vert\mathfrak{A}^j_{\mathcal{D}}\vert \le K^\prime\frac{w(j)^{V-I}(2j+1)^{2(F-B)}}{(j^2+\rho^2)^I}< \infty
\end{equation}
where the last inequality stems from \eqref{power-count} which has been shown to hold true.\\
One concludes that all the ribbon amplitudes stemming from \eqref{critical-action} are finite so that $S^f_{\Omega=1}$ is perturbatively finite to all orders.\par%

\section{Discussion}\label{section4}

Natural families of gauge invariant actions supporting a gauge invariant harmonic term can be constructed on $\mathbb{R}^3_\lambda$. This last property, which does not hold true on Moyal spaces, stems from the fact that the gauge invariant factor $\sim x_\mu x^\mu=x^2$ of the harmonic term, linked to the sum of the squares of the components of the gauge invariant canonical 1-form connection as defined in \eqref{component-invar-connect} belongs, actually to the non trivial center of the algebra $\mathbb{R}^3_\lambda$. Restricting ourselves to positive functional actions depending on the covariant coordinates (says $\Phi_\mu$ defined e.g by \eqref{tens-form}) which support a trivial global vacuum, a suitable BRST gauge-fixing gives rise to a family of matrix models with quartic interactions and kinetic operator with compact resolvant while the ghost sector decouples. The resulting functional action is given by $S_\Omega^f(\Phi)$ \eqref{quasilsz} where $\Omega$ is the real coefficient of $\{\Phi_\mu,\Phi_\nu\}^2$ involved in the classical gauge-invariant action.\\
Note that in the Moyal case, a harmonic term can be generated into the action as resulting from a gauge-fixing through the introduction of a suitable BRST-exact term \cite{blaschke-epl}. This yields a gauge propagator with the spectral properties needed to deal with the UV/IR mixing. Whether or not this interesting modification leads ultimately to a renormalisable gauge theorie on $\mathbb{R}^4_\theta$ remains to be seen.\par

We have considered the case $\Omega=1$ with 2 different types of traces, one being related to the canonical trace on $\mathbb{R}^3_\lambda$ and the other one reproducing the usual behavior of the Lebesgues integral on $\mathbb{R}^3$ as discussed in the subsection \ref{subsection21}. We have first computed the 2-point and 4-point functions at the 1-loop order and have found finite expressions. Perturbative finitude of all the amplitudes has been then extended to all orders. This perturbative finitude of $S^f_{\Omega=1}$ may be viewed as the result of the conjunction of 3 features:
\begin{enumerate}[i)]
 \item a sufficient rapid decay of the propagator at large indices (large $j$) so that correlations at large separation indices disappear,
\item the special role played by $j$, the radius of the fuzzy sphere components as a (UV/IR) cut-off, 
\item the existence of an upper bound for the (positive) propagator that depends only of the cut-off.
\end{enumerate}\par 

The above analysis can be extended to the case $\Omega\ne1$ for which the relevant action is given by \eqref{stot}-\eqref{squart}. The relevant kinetic operator is defined by%
\begin{eqnarray}
Q^{j_1j_2}_{mn;kl} &=& 8\pi\lambda^3w(j_1) \delta^{j_1j_2} \Lambda^{j_1}(k,l)\delta_{mn}\delta_{kl}\label{propaQ} \\
\Lambda^{j}(k,l)&=& M+\lambda^2\mu j(j+1)+\frac{\Omega}{2\lambda^2}(k+l)^2+\frac{4-3\Omega}{2\lambda^2}(k-l)^2 \ , \label{spectrumQ}
\end{eqnarray}
for any $j\in\frac{\mathbb{N}}{2},\ -j\le m,n,k,l\le j$. Note that the spectrum of $Q$ is positive, which is obvious from \eqref{spectrumQ}. The corresponding propagator is given by%
\begin{equation}
(Q^{-1})^{j_1j_2}_{mn;kl} = 
\frac{ \delta^{j_1j_2} \ \delta_{ml} \ \delta_{kn}}{8\pi\lambda^3 \ w(j_1) \ \left(M +\lambda^2\mu j_1(j_1+1) + \frac{\Omega}{2\lambda^2} (k+l)^2 + \frac{4-3\Omega}{2\lambda^2} (k-l)^2 \right) } \ . \label{q-1}
\end{equation}
As for the case $\Omega=1$ the propagator \eqref{q-1} verifies the following estimate%
\begin{equation}
0\le (Q^{-1})^{j_1j_2}_{mn;kl}\le(G^{-1})^{j_1j_2}_{mn;kl} \ , \quad \forall j_1,j_2\in\frac{\mathbb{N}}{2} \ , \quad -j\le m,n,k,l\le j \ . \label{envelop-number2}
\end{equation}
Thanks to this estimate, the analysis carried out above for the amplitudes of the $\Omega=1$ theory can be reproduced for $S^f_{\Omega\ne1}$ in a way similar to the one followed in the subsection \ref{subsection33} showing finitude of the corresponding amplitudes to all orders in perturbation. As a remark, we note that from the parameter dimensions \eqref{mass-dim} and the general expressions for the trace \eqref{trace-family} and kinetic terms $S_{Kin}\sim\frac{1}{g^2}\tr(\Phi K\Phi)$, the large $j$ (large indices) limit $j\to\infty$ can be interpreted naturally as the UV regime while $j=0$ corresponds to the IR regime. Hence, all the gauge theories on $\mathbb{R}^3_\lambda$ considered in this paper are UV finite with no IR singular behavior insured by condition \eqref{thecondition-positivity}.\par

The gauge theories considered here describe fluctuations of the covariant coordinate \eqref{tens-form} around the vacuum $\mathcal{A}^0_\mu=0$ (or alternatively the fluctuations of a gauge potential $A_\mu$ around the gauge potential $A^0_\mu=\theta_\mu$ defined by the gauge-invariant connection, in view of \eqref{tens-form}). The gauge theories considered in \cite{gervitwal-13} correspond to a choice $\mathcal{A}^0_\mu\ne0$ (or $A^0_\mu=0$). Then, expanding the classical gauge-invariant action $S(\mathcal{A})$ around this vacuum generates cubic interaction terms responsible for the occurrence of a non-zero tadpole showing up at the one-loop order leading to a vacuum instability. This is one major difference between the present work and \cite{gervitwal-13} (apart from more technical differences such as gauge choice and/or parameter choice). Note that the generic action for the family of gauge models in \cite{gervitwal-13} when truncated to a single fuzzy sphere component of the orthogonal sum in $\mathbb{R}^3_\lambda$ \eqref{orthogonal-decomp} is the action for the Alekseev-Recknagel-Schomerus model \cite{ARS} describing the low energy action for brane dynamics on $\mathbb{S}^3$. It would be interesting to see if a similar relation still exists with the family of gauge models considered here.\\
One aspect which deserves further study is to investigate carefully the commutative/semi-classical limit of the gauge theories considered in this paper and in \cite{gervitwal-13} in the spirit of what has been done e.g in \cite{jurman-steina}. Recall that the commutative limit of one of the traces considered here (the one for which $w(j)=j+1$) has been already investigated in \cite{pv-ksmap}, \cite{gervitwal-13} and formally shown to reproduce the usual Lebesgue integral on $\mathbb{R}^3$ while the fate of (gauge-fixed) kinetic operators that may occur in these gauge theories is not known so far.\par

As pointed out in the subsection\ref{subsection31}, the gauge-fixed action $S^f_{\Omega}$ bears some similarity with the so-called duality-covariant LSZ model \cite{LSZ}. In fact, one observes that $S^f_{\Omega=\frac{1}{3}}$ \eqref{quasilsz} coincides {\it{formally}} with one of the actions investigated in \cite{LSZ} leading to an exactly solvable model. Whenever $\Omega=\frac{1}{3}$, the quartic interaction potential in \eqref{quasilsz} depends only on the monomial $(\Phi^\dag\Phi)$ while the (positive) kinetic operator is somewhat different from the one of \cite{LSZ}. In fact, the partition function can be factorized in obvious notations as
\begin{equation}
Z(Q) =\prod_{j\in\frac{\mathbb{N}}{2}} Z_j(Q), \label{zq}
\end{equation}
with
\begin{equation}
Z_j(Q) =\int{\mathcal{D}} \Phi^j \mathcal{D} \ \Phi^{\dag j} \ \text{exp}\left(-\frac{w(j)}{g^2}\tr_j \left( 2 \left(\Phi^j Q^j\Phi^{\dag j}+\Phi^{\dag j} Q^j\Phi^j\right)+\frac{64}{3}\left(\Phi^j\Phi^{\dag j}\Phi^j\Phi^{\dag j}\right) \right)\right), \label{zqj}
\end{equation}
where%
\begin{equation}
\mathcal{D} \Phi^j \ \mathcal{D} \Phi^{\dag j} := \prod_{-j\le m,n\le j} \mathcal{D} \Phi^j_{mn} \mathcal{D} \Phi^{\dag j}_{mn} \ , \label{measure}
\end{equation}
and $Q^j$ is given by \eqref{propaQ}-\eqref{spectrumQ}, with however the weight $w(j)$ factored out from \eqref{spectrumQ} as it appears in front of the argument of the exponential and $\tr_j$ and the matrix $\Phi^j\in\mathbb{M}_{2j+1}(\mathbb{C})$ have been defined in \eqref{traceb}. By combining a singular value decomposition of $\Phi^j$ with the Harish-Chandra/Itzykson-Zuber measure formula, a standard computation  allows us to put any factor $Z_j(Q)$ under the form
\begin{equation}
Z_j(Q) = \frac{1}{\Delta^2(Q^j)} \ {\cal{N}}^j(g^2) \ \det_{-j\le m,n\le j} \left(f(\omega^j_m+\omega^j_n)\right) \ , \label{partitionfactor}
\end{equation}
where ${\cal{N}}^j(g^2)$ is a prefactor which is not essential here, $\Delta(Q^j)$ is the Vandermonde determinant associated with the matrix $Q^j$, $\omega^j_k$ are the eigenvalues of the real symmetric matrix defined by $\Lambda^j(m,n)$ \eqref{spectrumQ} and
\begin{equation}
f(z) = \sqrt{\frac{\pi g^2}{128w(j)}}\ \text{erfc}(z\sqrt{\frac{w(j)}{64g^2}})\ e^{z\frac{w(j)}{64g^2}}\label{functionspect}
\end{equation}
where
${\text{erfc}}$ is the complementary error function defined by%
\begin{equation}
\text{erfc}(z) = \frac{2}{\sqrt{\pi}} \int_z^\infty dx \ e^{-x^2} \ , \quad \forall z \in \mathbb{R}.
\end{equation}
The ratio of determinants appearing in the RHS of \eqref{partitionfactor} already signals that any $Z_j$ can be related to a $\tau$-function such as those occurring in integrable hierarchies. The relevant one here is the 2-Toda hierarchy. The complete analysis will be presented elsewhere \cite{solvab-15}.

\vspace*{40pt}\noindent\textbf{Acknowledgments}: This work is dedicated to the memory of Daniel Kastler. Discussions with Nicola Pinamonti are gratefully acknowledged. J.-C. W. thanks Michel Dubois-Violette for discussions on the role of canonical connections in noncommutative geometry and Patrizia Vitale for valuable discussions on algebraic structures related to $\mathbb{R}^3_\lambda$. T.~J. would like to thank LPT-Orsay and J.-C. Wallet for the kind hospitality during his visit which was funded by the French Government and Rudjer Boskovic Institute. T.~J. would like to thank Stjepan Meljanac for his encouragement and support during various stages of this work. The work by T.~J. has been partially  by Croatian Science Foundation under the project (IP-2014-09-9582).

\setcounter{section}{0}
\appendix
\section{Properties of the kinetic operators}\label{operators}
To simplify the notations, we drop the overall factor $8\pi\lambda^3$ in \eqref{kin-op1} and first assume $w(j)=1$. Let $L(a)$ denotes the left-multiplication operator by any element $a$ of $\mathbb{R}^3_\lambda$. Self-adjointness of the classical corresponding kinetic operator can be shown by using eqn.\eqref{kin-op1} to define the unbounded operator $G$ as%
\begin{equation}
G:= M \bbone + \mu L(x^2) \label{kinet-harmonic},
\end{equation}
an element of $\mathcal{L}(\mathcal{H})$, the space of linear operators acting on 
\begin{equation*}
\mathcal{H}=\text{span}\{v^j_{mn} \ , \ j\in\frac{\mathbb{N}}{2} \ , \ -j\le m,n\le \} 
\end{equation*}
with natural Hilbert product $\langle a,b \rangle = \tr(a^\dag b)$ defined in \eqref{traceb}. Obviously, $G$ is symmetric. By using \eqref{x0-commut} and \eqref{nat-fourier}, one infers%
\begin{equation}
x_0 = \lambda \ \sum_{j,m} \ j \ v^j_{mm} \ , \ \mbox{ and } \quad x^2 = \lambda^2 \ \sum_{j,m} \ j(j+1) \ v^j_{mm}.
\end{equation}
Therefore, 
\begin{equation}
L(x^2)=\lambda^2\sum_{j,m}j(j+1)L(v^j_{mm}),
\end{equation}
i.e $L(x^2)$ is a sum of orthogonal projectors, hence a sum of self-adjoint operators, says $L(v^j_{mm}):\mathbb{R}^3_\lambda\to\mathbb{M}_{2j+1}(\mathbb{C})$. This stems from $v^j_{mm}v^j_{mm}=v^j_{mm}$ (see \eqref{fusion}) and $\mathbb{R}^3_\lambda=\oplus_{j\in\frac{\mathbb{N}}{2}}\mathbb{M}_{2j+1}(\mathbb{C})$ \eqref{orthogonal-decomp}. One concludes that the classical kinetic operator $G$ is self-adjoint.\\
The positivity of $G$ can be realized from its spectrum given by%
\begin{equation}
\text{spec}(G)=\{ \lambda_j = M + \lambda^2 \mu j(j+1) >0 , \ \forall j \in \frac{\mathbb{N}}{2}\}.\label{spect-G}
\end{equation}
The corresponding $(2j+1)^2$-dimensional eigenspaces are
\begin{equation}
\mathcal{V}_{j} = \text{span}\{ v^j_{mn}, \ -j\le m,n\le j \},\label{eigensp-G}
\end{equation}
for any $j\in\frac{\mathbb{N}}{2}$. The extension of this analysis to arbitrary polynomial $w(j)$ is easily achieved by performing a simple rescaling at each step of the above discussion.\par 

As a remark, we note that $R_G(z)=(G-z\bbone)^{-1}$, $\forall z\notin\text{spec}(G)$, the resolvant operator of $G$, is compact. Indeed, pick $z=0$. Then, one easily realizes from $\text{spec}(G)$ that the operator $R_G(0)$ has decaying spectrum at $j\to\infty$, still with finite degeneracy for the eigenvalues at finite $j$. Hence $R_G(0)$ is compact which extends to $R_G(z),\ z\notin\text{spec}(G)$ by making use of the resolvant equation.\par%

A similar analysis holds for the gauge-fixed kinetic operator $K$ \eqref{operator-K} when $\Omega=1$. Its spectrum is easily found to be given by
\begin{equation}
\text{spec}(K) = \{ \rho_{j,p} = M + \mu \lambda^2 j(j+1) + \frac{8}{\lambda^2} p^2>0 \},\ \forall j \in \frac{\mathbb{N}}{2} , \ -j \le p \le j, \label{spec-K} 
\end{equation}
using 
\begin{equation*}
x_3^2 = \lambda^2 \sum_{j,m} m^2 v^j_{mm}.
\end{equation*}
The corresponding eigenspaces are
\begin{equation}
\mathcal{V}_{j,k} = \text{span}\{v^j_{pq},\ -j \le q \le j,\ \vert p\vert=k \},\ k=1,2,...,j\label{eigensp-K},
\end{equation}
for any $j \in \frac{\mathbb{N}}{2}$ and one has $\dim\mathcal{V}_{j,k\ne0}=2(2j+1)$, $\dim\mathcal{V}_{j,0}=2j+1$ together with the expected orthogonal decomposition $\mathcal{V}_j=\oplus_k\mathcal{V}_{j,k}$. \\
Self-adjointness of $K$ still holds since it can be written as a sum of orthogonal projectors, in view of the above expression for $x_3$ while positivity of $K$ is obvious from the spectrum \eqref{spec-K}.\par%

\section{Connected 2-point function at one-loop}\label{2-point-comput}

One starts from the relevant contribution of \eqref{connected-funct} to the connected 2-point function at one-loop written as
\begin{equation}
W(\mathcal{J}_\alpha)=W_0(\mathcal{J}_\alpha)-e^{-W_0(\mathcal{J}_\alpha)}S_4(\mathcal{J}_\alpha)e^{W_0(\mathcal{J}_\alpha)}+...\label{B-W}
\end{equation}
with
\begin{eqnarray}
S_4(\mathcal{J}_\alpha)&=&\sum\frac{32\pi\lambda^3}{g^2}w(j)\big((\frac{\delta}{\delta(\mathcal{J}_1)^j_{mn}}\frac{\delta}{\delta(\mathcal{J}_1)^j_{np}}\frac{\delta}{\delta(\mathcal{J}_1)^j_{pr}}\frac{\delta}{\delta(\mathcal{J}_1)^j_{rm}}+1\to2)\nonumber\\
&+&2(\frac{\delta}{\delta(\mathcal{J}_1)^j_{mn}}\frac{\delta}{\delta(\mathcal{J}_1)^j_{np}}\frac{\delta}{\delta(\mathcal{J}_2)^j_{pr}}\frac{\delta}{\delta(\mathcal{J}_2)^j_{rm}})\big).\label{B-S4}
\end{eqnarray}
To simplify the notations, it will be convenient to define 
\begin{equation}
(\mathcal{P}_\alpha)^j_{mn}:=-\frac{1}{2}P^j_{mn;kl}(\mathcal{J}_\alpha)^j_{kl},\ 
\alpha=1,2, -j\le m,n\le j\label{B-1stsol}
\end{equation}
for any $j\in\frac{\mathbb{N}}{2}$ which shows up naturally when using the Legendre transform to obtain the counterpart of \eqref{B-W} in the effective action $\Gamma(\Phi_\alpha)$. Indeed, one has
\begin{equation}
W(\mathcal{J}_\alpha)=-\Gamma(\Phi_\alpha)-\sum_{j,m,n}(\mathcal{J}_\alpha)^j_{mn}(\phi_\alpha)^j_{nm},\ \frac{\delta W(\mathcal{J}_\alpha)}{\delta(\mathcal{J}_\alpha)^j_{mn}}=-(\phi_\alpha)^j_{nm},\label{B-legendre}
\end{equation}
from which one realizes that the 1st order solution of the 2nd relation in \eqref{B-legendre}, which is needed in the present computation, is provided by \eqref{B-1stsol}, namely
\begin{equation}
\frac{\delta W_0(\mathcal{J}_\alpha)}{\delta(\mathcal{J}_\alpha)^j_{mn}}=-(\mathcal{P}_\alpha)^j_{mn}.
\end{equation}
After performing standard computation, we obtain (obvious summation indices not explicitely written)
\begin{eqnarray}
W(\mathcal{J}_\alpha))&=&\frac{1}{4}\sum(\mathcal{J}_\alpha)^{j_1}_{mn}P^{j_1j_2}_{mn;kl}(\mathcal{J}_\alpha)^{j_2}_{kl}-\frac{32\pi\lambda^3}{g^2}\sum w(j)(\mathcal{P}_1)^j_{mp}(\mathcal{P}_1)^j_{pr}(\mathcal{P}_1)^j_{rn}(\mathcal{P}_1)^j_{nm}\nonumber\\
&+&(1\to2)+2(\mathcal{P}_1)^j_{mp}(\mathcal{P}_1)^j_{pr}(\mathcal{P}_2)^j_{rn}(\mathcal{P}_2)^j_{nm})-\frac{32\pi\lambda^3}{g^2}\sum w(j)(P^j_{rm;pr}P^j_{np;mn}\nonumber\\
&+&\frac{1}{4}P^j_{rm;np}P^j_{pr;mn}
+[\frac{1}{2}P^j_{rm;mn} (\mathcal{P}_1)^j_{pr}(\mathcal{P}_1)^j_{np}+\frac{1}{2}P^j_{pr;mn} (\mathcal{P}_1)^j_{rm}(\mathcal{P}_1)^j_{np}\nonumber\\
&+&\frac{1}{2}P^j_{np;mn} (\mathcal{P}_1)^j_{pr}(\mathcal{P}_1)^j_{rm}+ \frac{1}{2}P^j_{rm;np} (\mathcal{P}_1)^j_{pr}(\mathcal{P}_1)^j_{mn}\nonumber\\
&+&\frac{1}{2}P^j_{pr;np} (\mathcal{P}_1)^j_{rm}(\mathcal{P}_1)^j_{mn}+\frac{1}{2}P^j_{rm;pr} (\mathcal{P}_1)^j_{np}(\mathcal{P}_1)^j_{mn}+P^j_{rm;pr} (\mathcal{P}_1)^j_{np}(\mathcal{P}_1)^j_{mn}\nonumber\\
&+&(1\to2)])+...\label{B-fullW}.
\end{eqnarray}
By making use of the Legendre transform \eqref{B-legendre} with
\begin{equation}
(\mathcal{J}_\alpha)^j_{sr}=-2K^j_{rs;nm}(\phi_\alpha)^j_{nm},\ \alpha=1,2\label{source-loworder},
\end{equation}
and taking into account the symmetries of the propagator stemming from \eqref{sym-K}, we finally obtain the expression for the relevant part of the effective action 
\begin{eqnarray}
\Gamma(\Phi_\alpha)&=&\sum(\phi_\alpha)^j_{mn}K^j_{mn;kl}(\phi_\alpha)^j_{kl}+\frac{32\pi\lambda^3}{g^2}\sum w(j)\tr_j((\Phi_1^2+\Phi_2^2)^2)\nonumber\\
&+&\frac{32\pi\lambda^3}{g^2}\sum w(j)\big((\phi_\alpha)^j_{pr}P^j_{rm;np}(\phi_\alpha)^j_{mn}+3(\phi_\alpha)^j_{pr}P^j_{rm;mn}(\phi_\alpha)^j_{np} \big)+...\label{B-Gamma}
\end{eqnarray}
in which the last two terms corresponds to the one-loop corrections.
\small
\section{Loop summation for the truncated model.}\label{appendix3}
Consider the loop built from from any $N$-point sub-diagram ${\cal{A}}_{m_1,n_1,...,m_N,n_N}$ and a propagator \eqref{g-1}. This latter can be taken to be $(G^{-1})^j_{m_1n_1;m_2n_2}$ without loss of generality. The corresponding $N-2$ amplitude is
\begin{equation}
\mathbb{A}_{m_3,n_3,...,m_N,n_N}=\sum_{-j\le m_1,n_1,m_2,n_2\le j}{\cal{A}}_{m_1,n_1,...,m_N,n_N}(G^{-1})^j_{m_1n_1;m_2n_2}\label{C-1},
\end{equation}
where the $N$-point part can be written generically as
\begin{equation}
{\cal{A}}_{m_1,n_1,...,m_N,n_N}=F_N(j)\prod_{p=1}^N\delta_{m_pn_{\sigma(p)}}\label{C-2}
\end{equation}
where $\sigma\in\mathfrak{S}_{N}$ is some permutation of $\{1,2,...,N \}$ and $F_N(j)$ is some function depending on $j$ and the other parameters of the model. Combining \eqref{C-1} with \eqref{C-2} and performing two summations, one obtains
\begin{equation}
\mathbb{A}_{m_3,n_3,...,m_N,n_N}=\frac{F_N(j)\Pi(J,M)}{w(j)}\sum_{-j\le n_1,n_2\le j}\big(\prod_{p=3}^N \delta_{m_pn_{\sigma(p)}}\big)\delta_{n_{\sigma(1)}n_2}\delta_{n_{\sigma(2)}n_1}\label{C-3}.
\end{equation}
The actual value of \eqref{C-3} is rules by the permutation $\sigma$. If $\sigma(1)=2$ and $\sigma(2)=1$ or $\sigma(1)=1$ and $\sigma(2)=2$ , \eqref{C-3} yields obviously
\begin{eqnarray}
\mathbb{A}_{m_3,n_3,...,m_N,n_N}&=&(2j+1)^2\frac{F_N(j)\Pi(J,M)}{w(j)}\big(\prod_{p=3}^N \delta_{m_pn_{\sigma(p)}}\big)\ {\text{when}}\ \sigma(1)=2,\ \sigma(2)=1\label{C-4},\\
\mathbb{A}_{m_3,n_3,...,m_N,n_N}&=&(2j+1)\frac{F_N(j)\Pi(J,M)}{w(j)}\big(\prod_{p=3}^N \delta_{m_pn_{\sigma(p)}}\big)\ {\text{when}}\ \sigma(1)=1,\ \sigma(2)=2\label{C-5}
\end{eqnarray}
thanks to the 2 last delta functions. When $\sigma(1)=2$ and $\sigma(2)\ne1$, the summation over $n_2$ can be readily performed to give
\begin{equation}
\mathbb{A}_{m_3,n_3,...,m_N,n_N}=(2j+1)\frac{F_N(j)\Pi(J,M)}{w(j)}\sum_{n_1=-j}^j\big(\prod_{p=3}^N \delta_{m_pn_{\sigma(p)}}\big)\delta_{n_{\sigma(2)}n_1}\label{C-6}.
\end{equation}
One further observes that $\sigma(2)$ is valued in $\{3,4,...,N \}$ so that there exists one $p_0\in\{3,...,N \}$ such that $\sigma(p_0)=1$. Hence, the last summation can be performed to give
\begin{equation}
\mathbb{A}_{m_3,n_3,...,m_N,n_N}=(2j+1)\frac{F_N(j)\Pi(J,M)}{w(j)}\big(\prod_{p=3, p\ne p_0}^N \delta_{m_{p_0}n_{\sigma(2)}}\big)\label{C-7},
\end{equation}
so that the loop summation produces an overall factor $(2j+1)$. A similar conclusion holds true for the case $\sigma(2)=1$ and $\sigma(1)\ne2$. Finally, in the remaining case $\sigma(1)\ne1,2$, $\sigma(2)\ne1,2$, the 2 summations simply yields a product of $N-2$ delta functions.

\end{document}